\documentclass[preprint]{aastex62}
\usepackage[utf8]{inputenc}
\usepackage{graphicx}
\usepackage{wrapfig}
\usepackage{array}
\usepackage{tabularx}
\usepackage{lineno}
\usepackage{amsmath, mathtools}
\usepackage{lineno}
\usepackage{xcolor}

\begin{document}


\title{Spectral trends across the rings and inner moons \\ of Uranus and Neptune from JWST NIRCam images}

\author{M.M. Hedman}
\author{I. de Pater}
\author{R. J. Cartwright}
\author{M. El Moutamid}
\author{R. DeColibus}
\author{M.R.  Showalter}
\author{M.S. Tiscareno}
\author{N. Rowe-Gurney}
\author{M.T. Roman}
\author{L N. Fletcher}
\author{H.B. Hammel}

\begin{abstract}
JWST NIRCam images provide low-resolution spectra of the rings and inner moons orbiting Uranus and Neptune. These data reveal systematic variations in spectral parameters like the strength of the strong OH absorption band around 3 $\mu$m and the spectral slopes at continuum wavelengths. Neptune's rings show an extremely weak 3-$\mu$m band, which is likely due to the small particle sizes in these dusty rings. Neptune's small inner moons also  have weaker 3-$\mu$m bands and redder continua than Uranus' small inner moons, indicating that Neptune's moons have a lower water-ice fraction. There are also clear spectral trends across the inner Uranian system. The strength of the 3-$\mu$m band clearly increases with distance from Uranus, with the rings having a noticeably weaker 3-$\mu$m band  than most of the small inner moons, which have a weaker 3-$\mu$m band than the larger moons like Miranda. While the rings and most of the small moons have neutral spectra between 1.4 $\mu$m and 2.1 $\mu$m, the outermost small moon Mab exhibits a blue spectral slope comparable to Miranda, indicating that Mab's surface may also be relatively water-ice rich. The next moon interior to Mab, Puck, exhibits a stronger 3-$\mu$m band and bluer continuum slope than any of the moons orbiting interior to it, perhaps indicating that it is being covered by water-ice-rich material derived from Mab via the $\mu$ ring. Finally, the small moon Rosalind has a redder spectral slope than its neighbors, possibly due to being coated with material from the dusty $\nu$ ring.
\end{abstract}

\section{Introduction}

\begin{figure}[h]
\resizebox{6.5in}{!}{\includegraphics{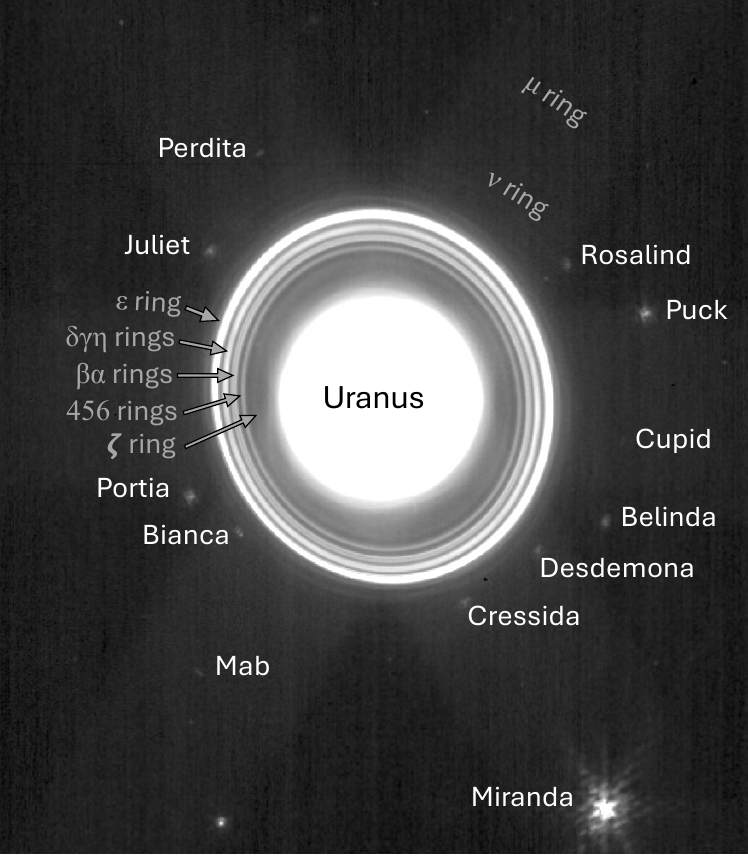}}
\caption{JWST NIRCam image showing Uranus' rings and small moons. This is a cropped version of image jw02739011001\_02105\_00003\_nrcb1\_i2d.fits, which was obtained using the F140M filter and calibrated using the standard pipeline \citep{Bushouse23}. All of the rings and moons considered in this paper are labeled and visible in this image. {Unmarked bright spots are either background stars or instrumental ariifacts.} Note the moons appear smeared due to their motion over the course of the integration. Also note that the dusty $\mu$ and $\nu$ rings are just barely visible in this image because it has been stretched to better show the moons.}
\label{Uranusim}
\end{figure}

\begin{figure}
\resizebox{6.5in}{!}{\includegraphics{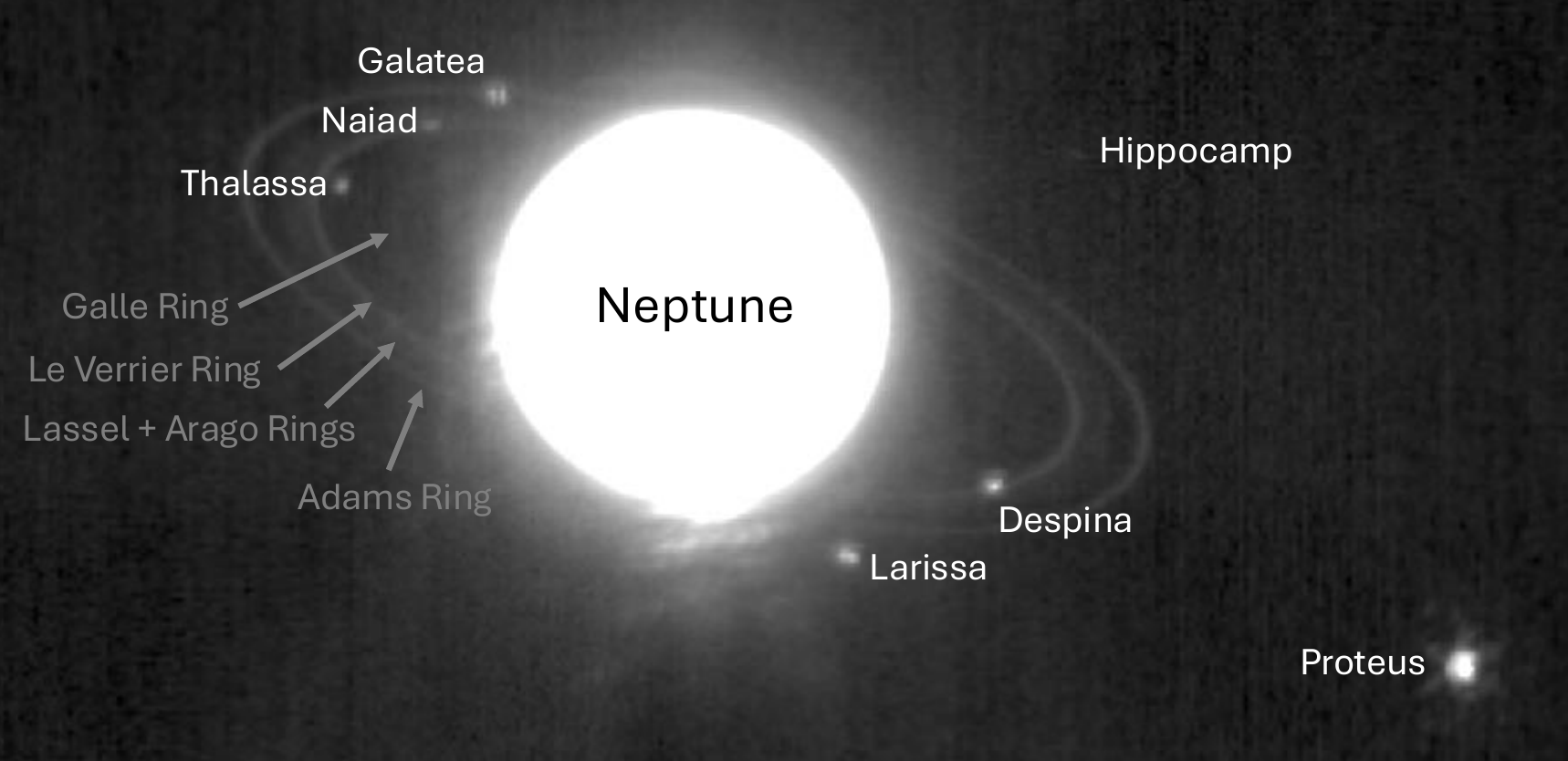}}
\caption{JWST NIRCam image showing Neptune's rings and small moons. This is a cropped version of image jw02739004001\_02105\_00003\_nrcb1\_i2d.fits which was obtained using the F140M filter and calibrated using the standard pipeline \citep{Bushouse23}. All of the rings and moons  considered in this paper are labeled and visible in this image. {Unmarked bright spots are either background stars or instrumental artifacts.} Note the moons appear smeared due to their motion over the course of the integration.}
\label{Neptuneim}
\end{figure}

Both Uranus and Neptune are surrounded by complex systems of rings and moons that are clearly visible in JWST images (see Figures~\ref{Uranusim} and~\ref{Neptuneim}). Despite its large obliquity, Uranus exhibits a reasonably regular ring-moon system, with a set of 9 narrow dense rings (designated 6 ,5 ,4, $\alpha$, $\beta$, $\eta$, $\gamma$, $\delta$ and $\epsilon$) located mostly\footnote{The exception is the innermost small moon Cordelia, which orbits interior to the outermost $\epsilon$ ring.} interior to a suite of thirteen small moons with radii ranging between 6 and 81 km (named Cordelia, Ophelia, Bianca, Cressida, Desdemona, Juliet, Portia, Rosalind, Cupid, Belinda, Perdita, Puck and Mab), outside of which are five larger satellites (Miranda, Ariel, Umbriel, Titania and Oberon) with radii between 230 km and  790 km \citep{Karkoschka01b, Thomas88, French91, Esposito91, Veverka91}. By contrast, the Neptune system is dominated by the single large moon Triton {(with a radius of 1353 km)}, whose retrograde orbit strongly suggests that it is a captured object \citep{McKinnon95, AH06, Nogueira11, RC17}. However, interior to Triton's orbit there is a set of dusty rings (named Galle, Le Verrier, Lassel, Arago and Adams) and seven moons {with radii between 17 km and 210 km} (Naiad, Thalassa, Despina, Galatea, Larissa, Hippocamp and Proteus) that are all on regular prograde orbits, implying that they are likely remnants of Neptune's original satellite system \citep{Thomas95, Porco95, Karkoschka03}. 

The spectra of these rings and moons encode important information about the origins and histories of the Uranus and Neptune systems. For example, high-quality spectra of  Triton show features due to methane, carbon dioxide, carbon monoxide and nitrogen ices, similar to Pluto \citep{Grundy03, Grundy10, Holler16, Wong23}, which helps to clarify the origin of that moon. Meanwhile, high-quality spectra of the larger Uranian moons show that they are all dominated by features due to water ice and carbon dioxide \citep{Bauer02, Grundy06, Cartwright18, Cartwright20b, Cartwright24}, and trends among these features are generally compatible with ongoing production and transport of volatiles within the Uranus system.  Spectral data for the smaller inner moons and rings are much more limited, but the available information already contains intriguing features that could provide additional insights into the origins and evolution of these systems. 

In particular, the rings and the small inner moons of Uranus have much lower reflectivities than any of the five large moons \citep{Karkoschka01, Karkoschka03, Gibbard05, Paradis19, Paradis23,  deKleer13, Nicholson18}.  Furthermore, recent studies of the JWST NIRCam images of selected small moons have revealed that these objects also have much weaker  3-$\mu$m water-ice absorption bands than the larger moons \citep{Belyakov24}. There are at least two possible interpretations for this trend that connect with the history and long-term evolution of this system.  

This trend could be due to differences in the long-term evolution of the various objects. The five outer moons are large enough that they could have undergone some amount of differentiation \citep{Castillo23}, which could leave their surfaces enhanced in lighter materials like water ice. By contrast, the inner moons and ring particles are so small that they are unlikely to have differentiated. Furthermore, most of these moons orbit inside Uranus' synchronous radius (where the moon's orbit period equals the planet's spin period), and so tidal forces should cause many of them to migrate in towards the planet, where they will eventually be torn apart by tidal forces to form massive rings that would then spread outwards until they are far enough from the planet to re-form into moons. Indeed, the material in the rings and small moons could have been torn apart and re-assembled multiple times over the Solar System's history \citep{HesselbrockMinton19}. If this idea  is correct, then the surfaces of the rings and small moons could be more representative of the bulk composition of the circumplanetary material. 

Alternatively, this trend could represent a primordial composition gradient  that might have been created by the same event that gave rise to Uranus' high obliquity. This tilt is often attributed to a giant impact \citep{BL10, Reinhardt20, Rogoszinski20}, but a challenge with this scenario is that an impact into the planet would not necessarily also re-orient the entire satellite system. To address this issue, it has been argued that the collision produced a massive compact debris disk around the planet (known as the `c-disk') that facilitated the re-alignment of the satellite disk (or `s-disk') by accelerating differential orbital precession \citep{Morbidelli12, Rufu22}.  In this case, the composition of the current rings and inner moons could reflect the composition of the original inner c-disk, while that of the larger moons would reflect that of the original satellite disk. 

Meanwhile, recent studies of the JWST NIRCam images of Neptune's small moons have shown that these objects also exhibit an absorption band around 3 $\mu$m \citep{Belyakov24}.  This may be compatible with the idea that Neptune's original satellite system was more similar in composition to Uranus' satellite system. However, the 3-$\mu$m feature also appears to be somewhat weaker on Neptune's moons than on Uranus' moons \citep{Belyakov24}, perhaps indicating a systematic difference in the compositions of the two satellite systems that could either be primordial or due to contamination from Triton's capture. 

In order to better understand these variations among these rings and moons, it is important to more completely document the trends among the various rings and small inner moons.  For example, Uranus' outermost small moon Mab exhibits a blue spectral slope that is more like Miranda than other small moons \citep{Molter23}.  This could potentially be evidence for a primordial compositional gradient across the Uranus system, since it indicates that the location of the moons (not just their size) affects their surface composition.  On the other hand, the innermost small moons all have sufficiently high densities to hold themselves together against tidal disruption \citep{French24}. {This provides evidence that tidal forces also  play} an important role in the long-term evolution of this system.  However, there are also ongoing processes that further complicate interpretation of these trends. In particular, Mab generates the dusty and spectrally blue $\mu$ ring that consists of fine particles that are transported inwards to Puck, while the dusty and spectrally red $\nu$ ring overlaps the orbits of Portia and Rosalind \citep{Showalter06, dePater06}. Untangling all of these processes requires spectral data on as many of the small moons as possible. 

JWST has obtained multiple high-quality images of the Uranus and Neptune system as part of DD program 2739 \citep{Pontoppidan22} and GTO program 2768  \citep{RoweGurney22}.  These images captured not only Uranus itself, but also all of its rings and small moons through as many as 7 different filters, providing broad-band spectral information for the rings and moons between 1 $\mu$m and 5$\mu$m. While a subset of these data have been used to quantify the spectral properties of a number of the small moons \citep{Belyakov24}, that earlier study did not consider the rings or several of the smaller moons that were captured in these images. We therefore perform a more comprehensive analysis of these images to obtain a relatively complete set of broad-band spectra for most of the rings and small moons orbiting Uranus and Neptune (Analyses of Uranus' faint $\mu$ and $\nu$ rings  are provided in a separate work by de Pater et al. 2025). This analysis also uses different methods for extracting the spectra of the moons that yield more consistent spectra, enabling trends within and between the Uranus and Neptune systems to be more clearly identified. 

Section~\ref{methods} below describes the relevant JWST images and how they were processed to obtain low-resolution spectra of the rings and small moons of both Uranus and Neptune. Section~\ref{results} describes the resulting spectra, while Section~\ref{discussion} discusses a number of interesting trends and variations among the spectra of these different objects. Finally, Section~\ref{Summary} provides a summary of our findings.

\section{Methods}
\label{methods}

\begin{table}
\caption{Geometric Parameters for Uranus and Neptune Observations}
\label{geomtab}
\hspace{-.5in}\resizebox{\textwidth}{!}{\begin{tabular}{|c|c|c|c|c|c|c|} \hline
Observation & Filters & Target & Ring Opening & Phase Angle & Planet-Observer & Planet-Sun \\
& & & Angle (deg) & (degrees) & Distance (AU) & Distance (AU) \\ \hline
DD2739-003 & F140M, F300M & Uranus & 56.3 & 2.90 & 19.67 & 19.66 \\
DD2739-011& F140M, F210M, F300M, F460M & Uranus & 64.1 & 2.81 & 19.28 & 19.63 \\
GTO2768-001& F182M, F210M, F410M, F480M & Uranus & 60.3 & 2.91 & 19.55 & 19.61 \\
GTO2768-002& F182M, F210M, F410M, F480M & Uranus & 60.3 & 2.91 & 19.55 & 19.61 \\
GTO2768-003& F182M, F210M, F410M, F480M & Uranus & 60.3 & 2.91 & 19.55 & 19.61 \\
DD2739-004 & F140M, F210M, F300M, F460M & Neptune & 21.5 & 1.79 & 29.47 & 29.92 \\
\hline
\end{tabular}}
\end{table}

This study examines NIRCam images of Uranus and Neptune that were obtained as part of DD program 2739 \citep{Pontoppidan22} and images of Uranus obtained for GTO program 2768 \citep{RoweGurney22}. {These data were obtained from the Mikulski Archive for Space Telescopes (MAST) at the Space Telescope Science Institute, and can be accessed via \dataset[doi: 10.17909/qtzb-3439]{https://doi.org/10.17909/qtzb-3439}.} Table~\ref{geomtab} summarizes these observations'  geometry, which were derived from Horizons via astroquery \citep{astroquery}. DD Program 2739 observed both Uranus and Neptune using NIRCam's F140M, F210M, F300M and F460M filters and two different observation modes. One observation used the RAPID mode with 5 integrations. Each integration had a different pointing and consisted of 8 groups, yielding an exposure duration of 85.9 seconds per pointing. The other observation used BRIGHT1 mode with 5 integrations. Again, each integration had a different pointing, but this time each integration consisted of 7 groups, yielding an exposure duration of 289.9 seconds per pointing (Figures~\ref{Uranusim} and~\ref{Neptuneim} are both examples of these longer-exposure images). While the complete observing sequence was done once for both planets, an earlier attempt to perform this observation at Uranus yielded 4 integrations in the RAPID mode with the F140M and F300M filters.  GTO program 2768 observed Uranus three times using the F182M, F210M, F410M and F480M filters. Each observation used the  RAPID mode and had 5 integrations with 5 different pointings. Each of these integrations used 10 groups, and so the exposure duration was 107.4 seconds per pointing.

For this analysis, we use images that have been calibrated and reprojected into sky coordinates using the JWST calibration pipeline \citep{Bushouse23}. Note that calibrated and reprojected images were constructed for each individual pointing and for all the data obtained with a particular filter in each observation, and both these different types of calibrated images are used in this study. These images are in units of MJy/sr, but since the observed signals from the rings and small moons consist of reflected sunlight,  it is more useful to express these signals in terms of standardized measures of reflectance that depend upon the effective solar flux for each NIRCam filter. Table~\ref{filtertab} provides the effective average solar flux at 1 AU for each of the relevant NIRCam filters, which are computed by convolving the standard solar spectrum \citep{Rieke08}\footnote{Available at {https://www.stsci.edu/hst/instrumentation/reference-data-for-calibration-and-tools/astronomical-catalogs/solar-system-objects-spectra}} with the appropriate filter transmission functions\footnote{Available at {https://jwst-docs.stsci.edu/jwst-near-infrared-camera/nircam-instrumentation/nircam-filters}}. Similar convolutions were used to determine the effective average observed wavelength for each filter assuming the source has a sun-like spectrum, which are also provided in Table~\ref{filtertab}.

The procedures for obtaining spectra of the rings are very different from those used to obtain spectra of the small moons, so each set of procedures will be described separately below. {All the resulting spectra for the rings and moons show comparable features, and so the shapes of these spectra will be documented in the following Section~\ref{results}.}

\subsection{Obtaining Spectra of the Rings}

\begin{table}
\caption{Relevant Parameters for NIRCam Filter Calibration}
\label{filtertab}
\hspace{0.in}{\begin{tabular}{|c|c|c|c|c|} \hline
Filter & Wavelength & Effective Central &   Solar Flux$^b$ & PSF Correction  \\
Name &  Range$^a$ ($\mu$m) & Wavelength$^b$ ($\mu$m)  & (W/m$^2$/$\mu$m) & Factor$^c$  \\
\hline  
F140M & 1.331-1.479 & 1.403 & 359.465 & 0.616 \\
F182M & 1.722-1.968 & 1.837 & 140.465 & 0.632 \\
F210M & 1.992-2.201 & 2.090 & 102.312 & 0.644 \\
F300M & 2.831-3.157 & 2.986 & 26.470 & 0.678 \\
F410M & 3.866-4.302 & 4.069 & 8.127 & 0.667 \\
F460M & 4.515-4.748 & 4.627 & 4.829 & 0.667 \\
F480M & 4.662-4.963 & 4.808 & 4.190 & 0.687 \\
\hline
\end{tabular}}

$^a${\footnotesize From  {https://jwst-docs.stsci.edu/jwst-near-infrared-camera/nircam-instrumentation/nircam-filters}}

$^b${\footnotesize Value at 1 AU, computed using the standard solar spectrum \citep{Rieke08} and filter transmission functions.}

$^c${\footnotesize Computed based on the ratio of the gaussian fit signal from the star P330E in NIRCam images \cite{Gordon19} to the predicted flux derived from the CALSPEC spectrum. See Section 2.2 for details.}
\end{table}

\begin{table}
\caption{Ring Spectral Parameters}
\label{ringtab}
\resizebox{6.5in}{!}{\begin{tabular}{|c|c|c|c|}\hline
Filename & NEW (m) & NEW (m) & NEW (m) \\
 (Uranus) & Main Rings & $\epsilon$ & $6-\delta$\\
\hline
jw02739-o011\_t002\_nircam\_clear-f140m\_i2d.fits & 
1497.8$\pm$2.4 & 1117.6$\pm$2.2 & 376.5$\pm$1.4 \\
jw02739-o011\_t002\_nircam\_clear-f210m\_i2d.fits & 
1504.6$\pm$1.7 & 1153.4$\pm$1.5 & 339.5$\pm$1.0 \\
jw02739-o011\_t002\_nircam\_clear-f300m\_i2d.fits &
588.1$\pm$1.0 & 438.4$\pm$0.9 & 136.9$\pm$0.6 \\
jw02739-o011\_t002\_nircam\_clear-f460m\_i2d.fits & 
1387.7$\pm$2.2 & 1045.5$\pm$2.0 & 321.0$\pm$1.3 \\
\hline
jw02768-o001\_t001\_nircam\_clear-f182m\_i2d.fits & 
1421.5$\pm$2.0 & 1054.7$\pm$1.8 & 356.0$\pm$1.2 \\
jw02768-o002\_t001\_nircam\_clear-f182m\_i2d.fits & 
1424.4$\pm$ 2.0 & 1049.2$\pm$1.8 & 352.3$\pm$1.2 \\
jw02768-o003\_t001\_nircam\_clear-f182m\_i2d.fits & 
1410.6$\pm$2.0 & 1041.7$\pm$1.7 & 351.4$\pm$1.2 \\
jw02768-o001\_t001\_nircam\_clear-f210m\_i2d.fits & 
1440.4$\pm$1.7 & 1094.4$\pm$1.5 & 338.3$\pm$1.0\\
jw02768-o002\_t001\_nircam\_clear-f210m\_i2d.fits & 
1441.5$\pm$1.7 & 1092.7$\pm$1.4 & 332.1$\pm$0.9 \\
jw02768-o003\_t001\_nircam\_clear-f210m\_i2d.fits & 
1445.1$\pm$1.5 & 1101.9$\pm$1.3 & 334.0$\pm$0.9 \\
jw02768-o001\_t001\_nircam\_clear-f410m\_i2d.fits & 
1305.5$\pm$1.4 & 977.9$\pm$1.3 & 308.4$\pm$0.9 \\
jw02768-o002\_t001\_nircam\_clear-f410m\_i2d.fits & 
1299.7$\pm$1.4 & 978.7$\pm$1.2 & 301.0$\pm$0.9 \\
jw02768-o003\_t001\_nircam\_clear-f410m\_i2d.fits & 
1292.0$\pm$1.4 & 973.2$\pm$1.3 & 298.3$\pm$0.9 \\
jw02768-o001\_t001\_nircam\_clear-f480m\_i2d.fits & 
1400.8$\pm$2.0 & 1049.9$\pm$1.8 & 325.7$\pm$1.2 \\
jw02768-o002\_t001\_nircam\_clear-f480m\_i2d.fits & 
1421.1$\pm$1.9 & 1074.2$\pm$1.7 & 326.8$\pm$1.2 \\
jw02768-o003\_t001\_nircam\_clear-f480m\_i2d.fits &
1411.8$\pm$1.9 & 1064.6$\pm$1.7 & 326.2$\pm$1.1 \\
\hline
F140M Average & 1497.7$\pm$ 2.4 & 1117.6$\pm$2.2 & 376.5$\pm$1.5 \\ 
F182M Average & 1418.9$\pm$4.2 & 1048.5$\pm$3.8 & 353.2$\pm$1.4 \\
F210M Average & 1457.9$\pm$15.5& 1110.6$\pm$14.4& 336.0$\pm$1.7 \\
F300M Average & 588.1$\pm$1.0& 438.4$\pm$0.9& 136.9$\pm$0.6\\
F410M Average & 1299.1$\pm$3.9& 976.6$\pm$1.7&302.6 $\pm$3.0\\
F460M Average & 1387.7$\pm$2.2& 1045.5$\pm$2.0& 321.0$\pm$1.3\\
F480M Average & 1411.2$\pm$5.9& 1062.9$\pm$7.1& 326.2$\pm$0.3\\
\hline
Filename &  NEW  (m) & NEW (m) & NEW (m) \\
(Neptune) &  Main Rings & Adams & Le Verrier+ \\
\hline
jw02739-o004\_t003\_nircam\_clear-f140m\_i2d.fits &
40.63$\pm$1.01 & 13.16$\pm$1.01 & 29.99$\pm$0.82 \\
jw02739-o004\_t003\_nircam\_clear-f210m\_i2d.fits & 
54.74$\pm$0.93 & 18.22$\pm$0.93 & 41.04$\pm$0.75\\
jw02739-o004\_t003\_nircam\_clear-f300m\_i2d.fits & 
46.43$\pm$0.42 & 15.91$\pm$0.42 & 34.97$\pm$0.34 \\
jw02739-o004\_t003\_nircam\_clear-f460m\_i2d.fits & 
61.74$\pm$1.70 & 21.88$\pm$1.69 & 45.63$\pm$1.36 \\
\hline
\end{tabular}}
\end{table}

The rings of Uranus and Neptune exhibit significant longitudinal brightness variations, but for this particular study we are only interested in the average spectral properties of the densest rings, so we can use the images constructed from all the exposures obtained with a given filter, which have the highest signal-to-noise. We therefore generate average radial brightness profiles of the rings from the level 3 i2d images from both DD observation 2739 and GTO observation 2768 listed in Table~\ref{ringtab}.\footnote{ Note we did not attempt to generate profiles from the first attempt to observe Uranus in DD observation 2739 both because the signal-to-noise of the observations were lower and because a bright cloud feature on Uranus produced significant background signals across a portion of the main rings.}

Radial brightness profiles of the rings were generated by computing the radius and longitude of each pixel in the planet's ring-plane based on a combination of the sky-plane coordinates of each pixel and the contemporary location and orientation of the planet provided by Horizons via astroquery \citep{astroquery}. We verified and refined this geometry based on the predicted location of the known rings by applying small offsets and rotations. After navigating each image, the imaging data are divided into 36 wedges, each of which corresponding to 10$^\circ$ in ring-plane longitude, and the brightness data in each wedge are sorted and interpolated onto a regular grid of radius values. For Uranus the planet is near solstice, which means that its rings are so open that they can be clearly seen all around the planet. We therefore take the average of all 36 profiles to obtain an estimate of the rings' average brightness as a function of distance from the planet. For Neptune, the rings are much less open, so we instead just average the brightness of the 6 profiles consisting of data within $\pm 45^\circ$ of each ansa (Note that this also excluded the bright arcs in the Adams ring). These profiles were then converted from units of MJy/sr to  a standard measure of ring reflectance called normal $I/F$ or $\mu I/F$, where $\mu$ is the sine of the ring opening angle,  $I$ is the intensity of the scattered light and $F$ is the solar flux density (flux divided by pi) at the rings.  Table~\ref{filtertab} provides the computed values for the solar flux at 1 AU and average observed wavelength for each of the relevant NIRCam filters, while Table~\ref{geomtab} provides estimates  of distance between the planets and the Sun, as well as the ring opening angle to JWST for each image using the same Horizons data that provided the overall image geometry.

All of the brightness profiles exhibited background trends due to signals from the planet itself. We  removed these trends by fitting the log-transformed data on either side of the main rings to a third-order polynomial and removing this trend from the observations. For Uranus, the background fit regions were 35,000-40,000 km and 58,000-60,000 km, while for  Neptune the background fit regions were  40,000-47,000 km and 67,000-80,000 km.  Note that these regions include the faint  $\zeta$ ring at Uranus and the Galle ring at Neptune, so these fits may slightly overcorrect for any faint dust sheets around the brighter main rings. However, this overcorrection should have minimal effect on the relatively bright rings that are the focus of this study.

\begin{figure}
\resizebox{\textwidth}{!}{\includegraphics{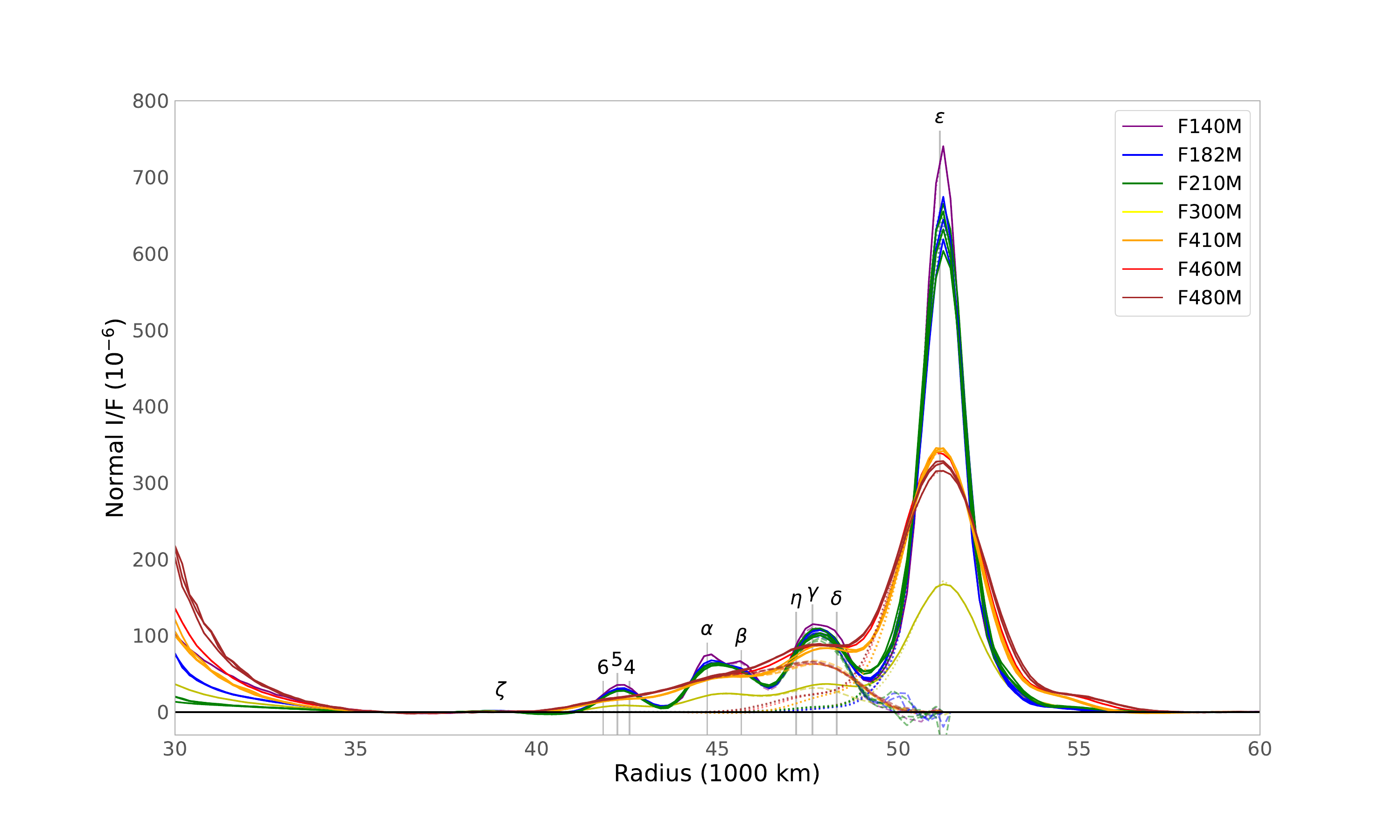}}
\caption{Average background-subtracted brightness profiles of Uranus' main rings derived from the NIRCam images. Different colored profiles correspond to data from different NIRCam filters, and the locations of the nine main rings and the dusty $\zeta$ ring are marked. The faint dotted and dashed lines show the versions of the profiles used to estimate the brightnesses of the $\epsilon$ and 6-$\delta$ rings, respectively. {Note that these separate profiles deviate from the observed profiles most noticeably in the transition between the $\epsilon$ and $\delta$ rings.} }
\label{uringprof}
\end{figure}

\begin{figure}
\resizebox{\textwidth}{!}{\includegraphics{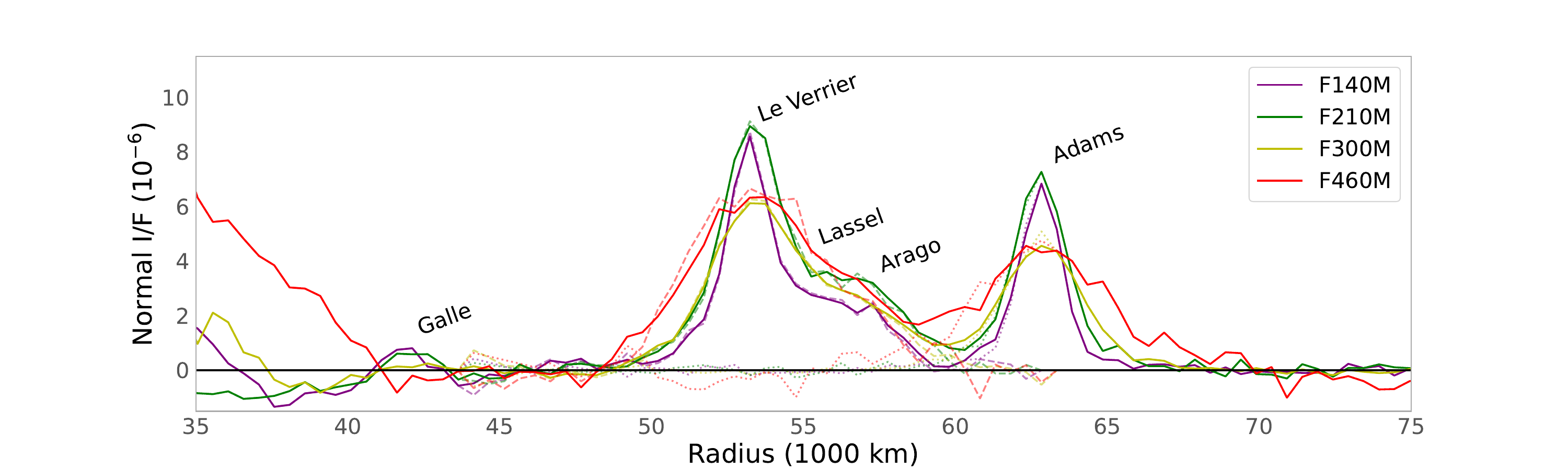}}
\caption{Average background-subtracted  brightness profiles of Neptune's rings derived from the NIRCam images. Different colored profiles correspond to data from the different NIRCam filters, and the various rings are marked. The faint dotted and dashed lines show the versions of the profiles used to estimate the brightnesses of the Adams ring and the combined Le Verrier, Lassel and Arago rings, respectively. {Note that these separate profiles deviate from the observed profiles most noticeably in the transition between the Adams and Arago rings.}}
\label{nringprof}
\end{figure}

Figures~\ref{uringprof} and~\ref{nringprof} show the average, background-subtracted brightness profiles derived from each of the different NIRCam images.  On the left side of each plot there is still a residual trend due to the planet, but the signal is clearly close to zero on either side of the relevant rings. It is also clear that the signal from Uranus' rings is much stronger than the signal from Neptune's rings, which is reasonable because Uranus' rings are dense rings with optical depths near unity, while Neptune's rings are much more tenuous and dusty \citep{dePater18, Nicholson18}. 

For all the Uranian ring profiles shown in Figure~\ref{uringprof}, the $\epsilon$ ring is the most prominent feature. In the profiles derived from the shorter-wavelength filters, one can also discern three bumps corresponding to three different sets of rings (456, $\alpha\beta$ and $\eta\gamma\delta$). The signals from these three groups of rings become more difficult to distinguish from each other at longer wavelengths, but the combined signal from all these rings is still distinguishable from the $\epsilon$ ring signal even for the longest-wavelength F480M filter.  The faint $\zeta$ ring is even visible in these profiles, but will not be considered further here because isolating this ring from the inner main rings requires special processing that is beyond the scope of this work. Similarly, the faint $\mu$ and $\nu$ rings found further out (see Figure~\ref{Uranusim}) are also detectable in these profiles, but {extracting} these weak signals requires additional processing that {is} discussed in a separate paper (de Pater et al. 2025).

The Neptune ring profiles shown in Figure~\ref{nringprof} also show consistent ring features, with the narrow Le Verrier and Adams rings appearing as well-defined peaks, and the Lassel and Arago rings appearing as a shelf of material extending outwards from the Le Verrier ring. The Galle ring is also visible in the shorter-wavelength profiles, but again we will not consider this ring further here because this faint ring requires more in-depth analysis to reliably quantify. 
 
Directly comparing the signals in these different profiles is complicated by their differing spatial resolutions. Fortunately, we can extract resolution-independent estimates of the rings' total brightness from these profiles by computing a quantity called the Normal Equivalent Width or NEW. In this context, NEW is simply the radially integrated $I/F$ above background over the region containing the ring. To obtain the average spectra of the Uranian main rings, we performed this integral over the radius range of 35,000-60,000 km, while for Neptune's main rings we performed this integral over the radius range of 45,000-70,000 km. The resulting NEW values {derived from each NIRCam image} are provided in Table~\ref{ringtab}. The error bars on these points are statistical 1-$\sigma$ error bars based on the standard deviation of the data points in the blank regions on either sides of the ring after background subtraction. Note that the scatter among the values derived from different images of Uranus using the same filters are larger than these uncertainties, implying that there are additional systematic errors in these brightness estimates. Even so, the scatter among these NEW values are still typically only of order a few percent, implying that the uncertainties in all of these estimates are better than 5\%.\footnote{Note that the absolute calibration of the various NIRCam filters is currently better than 4\% https://jwst-docs.stsci.edu/jwst-calibration-status/nircam-calibration-status/nircam-imaging-calibration-status\#NIRCamImagingCalibrationStatus-Photometriccalibration}. {In order to compare the Uranian ring spectra with those of the other rings and moons, we also compute the average NEW values for all measurements at the same wavelength. To account for the excess scatter among the observations, we estimated the uncertainty on each NEW value derived from multiple observations as the standard error on  the mean of the individual estimates.}

We did perform an initial investigation of these profiles to search for any evidence of spectral trends within the rings by either integrating over a sub-set of the radial ranges, or by fitting the profiles to appropriate model convolutions. However, we found that any spectral trends were rather subtle, and so for this analysis we will only isolate the signal from the outermost ring from each planet (the $\epsilon$ ring at Uranus and the Adams ring at Neptune) from the rest of the rings. {Despite these rings being the most distinct in the various profiles, close inspection of Figures~\ref{uringprof} and~\ref{nringprof} shows that the signals from each of these rings still overlaps the signals from other rings. In order to isolate the signals from each of these rings, we first fit a Gaussian to each profile to identify the peak location for the $\epsilon$ or Adams ring. We then reflect the brightness profile of the portion of the ring exterior to this peak location to estimate the brightness profile for the isolated ring interior to the peak location.  These profiles are illustrated by the dotted lines in Figures~\ref{uringprof} and~\ref{nringprof}, which are only visible interior to the peak of either the $\epsilon$ or Adams ring because they exactly overlap the data exterior to the peaks of these rings. Note that for the $\epsilon$ ring profiles shown in Figure~\ref{uringprof}, these profiles are clearly not simple Gaussians. Instead they have more complex shapes that represent the convolution of the NIRCam point-spread functions with the narrow ring. On the other hand, the Adams ring profiles deviate from zero more noticeably under the Le Verrier, Lassel and Arago rings in Figure~\ref{nringprof} due to the lower signal-to-noise in these profiles. Removing the estimated signals from the $\epsilon$ and Adams ring yields the profiles illustrated with dashed lines in Figures~\ref{uringprof} and~\ref{nringprof}, which represent the combined signals from the  $\delta$, $\gamma$, $\eta$, $\beta$, $\alpha$, 4, 5 and 6 rings at Uranus, or the Le Verrier, Arago and Lassel rings at Neptune. Note that all these profiles become zero exterior to the peak of the $\epsilon$ or Adams ring by construction, and converge towards the observed profile at sufficiently small radii.}

We compute the NEW of these different rings by integrating the appropriate profiles over 40,000-60,000 km for the $\epsilon$ ring, 40,000-49,000 km for the $6-\delta$ rings, 50,000-75,000 km for the Adams ring, and 46,000-62,000 km for the Le Verrier, Lassel and Arago rings. The resulting NEW values are provided in Table~\ref{ringtab}, along with 1-$\sigma$ statistical error bars computed based on the $rms$ variations in the blank regions around the entire ring system. {For the Uranian rings we also provide the average NEW values observed through each filter using the same techniques described above.}

\subsection{Obtaining Spectra of the small moons}

Spectra of selected small moons have already been extracted from the NIRCam images obtained as part of DD Program 2739 by \citet{Belyakov24}. For this analysis we wanted to extend this work to also consider the data from GTO Program 2768 and to obtain as much spectral information as possible about all the inner small moons of Uranus and Neptune. To achieve these goals, we followed procedures that deviate from those described in \citet{Belyakov24} in various ways that are highlighted at appropriate points below.

Since the small moons move substantially relative to their host planet over the course of each observation, it is inappropriate to use the same co-added images that we used above to obtain high signal-to-noise spectra of the rings. Indeed, the signals from the various moons are largely removed from those images by the standard calibration pipeline. We therefore instead use the calibrated versions of the images derived from individual pointings, each of which corresponds to a much shorter integration time. However, even the calibrated images constructed from a single pointing can pose problems for quantifying signals from small moons.  \citet{Belyakov24} already noted that in the longer exposure BRIGHT1 images from the DD program 2739 observation, the sudden rises and falls in a pixel's brightness as a moon moved into or out of it could cause issues with determining the count rate used in the standard calibration pipeline. 

We not only confirmed that the calibration pipeline impacted the appearance of moons in the BRIGHT1 images, but also found that the calibration pipeline has more subtle but measurable effects on the flux estimates in the other calibrated images. We discovered this broader issue when we compared estimates of selected moon's reflectances derived from the F182M and F210M images obtained as part of GTO program 2768. Specifically, we found that the F210M reflectances were always systematically higher than the F182M reflectances by a factor of order 20\%, which is inconsistent with the  expected spectral trends over this relatively limited wavelength range. This discrepancy could not be resolved by considering different techniques to correct for the effects of the point spread function. However, we were able to obtain consistent results at these two wavelengths by re-calibrating the data and deliberately skipping the `jump' step in stage one of the calibration pipeline, which is the one that checks whether the data number (DN) in each pixel increases linearly with time and excludes any data containing sufficiently large sudden jumps in brightness \citep{Bushouse23}. Removing this step also results in the images containing many more cosmic rays. While this could potentially be reduced by re-introducing the jump step with a different threshold, we found that the cosmic rays avoided the moons in enough of the images that such complications were not needed for this particular analysis. Also note that by skipping the jump step, we could obtain useful brightness estimates of the moons from the DD 2739 BRIGHT1 images. This was particularly useful for obtaining spectral data on a few of Neptune's fainter moons. 

Our version of the calibration pipeline also included the background subtraction step that {was not conducted by} { \citet{Belyakov24}  because we used different procedures for dealing with backgrounds due to the rings. We agree with \citet{Belyakov24} that the signals from the rings can impact the spectra of the innermost small moons, particularly for Neptune. However, instead of using theoretical models of the rings scaled to the features in each image, we instead used the radial brightness profiles derived in the previous subsection to estimate the ring background for all images obtained with the appropriate filters. 

For the Neptune images, we simply took the background-subtracted brightness profiles and interpolated these onto the radius values of each pixel in each image to generate a predicted image of the rings, which we then smoothed using a gaussian filter with a scale length of 1 pixel in order to match the projected resolution of the true image. Subtracting this model of the rings from the relevant image removed most of the ring signals in the vicinity of the moons, although it did leave some backgrounds from the planet itself. Note that this background subtraction was only applied to pixels where the predicted ringplane radius was less than 80,000 km, so this background subtraction step did not affect the regions around the outer moons Proteus and Hippocamp. 

By contrast, for the Uranus images the most concerning background is from the $\epsilon$ ring, whose brightness and location varies with longitude around the planet. We therefore instead took each of the profiles derived from a 10$^\circ$-wide region in the images listed in Table~\ref{ringtab} and interpolated each of those profiles into the appropriate wedge of each calibrated image to generate a model ring image. These model ring images were then subtracted from the real images without applying a gaussian filter to the former because we were hoping to be able to isolate the signals from the moons Cordelia and Ophelia, which are both very close to the $\epsilon$ ring.  The background subtraction did allow the signals from the innermost moons Cordelia and Ophelia to be clearly seen in the shorter-wavelength images, but we were unable to get reliable brightness estimates of these two moons. This background subtraction step also had slight effects on the spectra of Bianca, Cressida and Desdemona. For the sake of completeness we report brightness estimates derived both with and without this ring background subtraction for the Uranian moons out through Rosalind (see Tables~\ref{moontab1} and~\ref{moontab2}). 

We experimented with a number of different techniques for extracting the brightness of individual moons from the background-subtracted images. We opted not to align and co-add data from multiple images together to improve signal-to-noise like \citet{Belyakov24} did. {This was because we considered images of the moons that were all clearly smeared, and we did not want to combine data from images where the moon signals would have different shapes.} We therefore instead opted to extract separate estimates of the moons'  brightness from each individual image. This also had the advantage that it enabled uncertainties in these brightness estimates to be directly computed based on the scatter among the values derived from different images. However, this also meant that the signal-to-noise for many of the fainter moons was too low for simple aperture-integrated photometry to provide sufficiently stable estimates of the moons' brightness. We therefore instead estimated the signal from each moon in each image by fitting the data within a $9\times9$ pixel region centered on the moon to a two-dimensional Gaussian plus a background with constant and linear components, and then computing the integrated volume under the Gaussian. To account for the possibility of smear in the image, we allowed the widths and orientation of the Gaussian to float in these fits. 

These Gaussian fits yielded much more consistent estimates of each moon's brightness at each wavelength. However, these numbers still need to be corrected to account for the fact that the NIRCam point spread function is not a perfect Gaussian. This was accomplished by following a procedure analogous to that used by \citet{Belyakov24}, which employed observations of the star P330E, which has a well-calibrated near-infrared flux spectrum provided as part of the CALSPEC program \citep{Bohlin14, Bohlin15, Bohlin20}. We used the spectrum provided in the file {\tt p330e\_stiswfcnic\_007.fits} and convolved it with the appropriate NIRCam filter transmission functions in order to estimate the total flux from this star that should be measured by each NIRCAM filter configuration. We then took the calibrated NIRCam images of P330E obtained as part of JWST calibration program 1538 \citep{Gordon19} and fit $9\times9$ pixel regions to a gaussian using the same algorithm used with the various moons. The ratio of the flux derived from the Gaussian fit to the predicted total flux from the CALSPEC spectrum provides the correction factors listed in Table~\ref{filtertab} needed to convert our measured signals from the moons to proper estimates of the moons' total flux in each wavelength band. Note that these factors are all around 0.65, whereas the factors used by \citet{Belyakov24} were around 0.83-0.86. This difference is due to the fact that \citet{Belyakov24} used a direct integration of a wider aperture (radius of 6 pixels), which captured a larger fraction of the total flux from the moon.

\begin{table}
\caption{Average  fluxes of the observed moons (all in microJanskies, no corrections for variations in the distances to the planets included). {For the inner Uranian moons, the first row is without removal of the ring background and the second row is after removing the ring background.}}
\label{moontab1}
\hspace{-.5in}\resizebox{\textwidth}{!}{\begin{tabular}{|c|c|c|c|c|c|c|c|}
\hline
Moon & F140M & F182M & F210M & F300M & F410M & F460M & F480M \\ 
\hline
Bianca & 3.406$\pm$0.095 & 2.641$\pm$0.047 & 2.194$\pm$0.083 & 0.254$\pm$0.053 & 0.588$\pm$0.055 & 0.355$\pm$0.061 & 0.443$\pm$0.069\\
(ring bg removed) & 3.421$\pm$0.083 & 2.682$\pm$0.041 & 2.182$\pm$0.075 & 0.347$\pm$0.054 & 0.597$\pm$0.023 & 0.428$\pm$0.023 & 0.526$\pm$0.077\\ \hline
Cressida & 7.780$\pm$0.250 & 6.045$\pm$0.079 & 4.830$\pm$0.125 & 0.600$\pm$0.053 & 1.073$\pm$0.021 & 1.040$\pm$0.113 & 0.852$\pm$0.028\\
(ring bg removed) & 7.670$\pm$0.249 & 5.973$\pm$0.090 & 4.897$\pm$0.115 & 0.616$\pm$0.06 & 1.132$\pm$0.031 & 1.111$\pm$0.134 & 0.882$\pm$0.038\\ \hline
Desdemona & 6.378$\pm$0.238 & 4.846$\pm$0.092 & 3.844$\pm$0.110 & 0.464$\pm$0.009 & 0.896$\pm$0.027 & 0.764$\pm$0.039 & 0.706$\pm$0.051\\
(ring bg removed) & 6.299$\pm$0.236 & 4.862$\pm$0.081 & 3.817$\pm$0.147 & 0.473$\pm$0.018 & 0.932$\pm$0.030 & 0.756$\pm$0.072 & 0.698$\pm$0.060\\ \hline
Juliet & 13.389$\pm$0.712 & 10.619$\pm$0.098 & 8.540$\pm$0.214 & 0.983$\pm$0.053 & 1.969$\pm$0.027 & 1.606$\pm$0.049 & 1.496$\pm$0.053\\
(ring bg removed) & 13.191$\pm$0.702 & 10.507$\pm$0.080 & 8.632$\pm$0.216 & 0.986$\pm$0.051 & 1.980$\pm$0.029 & 1.646$\pm$0.043 & 1.607$\pm$0.069\\ \hline
Portia & 23.701$\pm$0.603 & 18.209$\pm$0.209 & 14.524$\pm$0.353 & 1.78$\pm$0.057 & 3.616$\pm$0.038 & 2.914$\pm$0.145 & 2.717$\pm$0.029\\
(ring bg removed) & 23.379$\pm$0.596 & 18.145$\pm$0.226 & 14.652$\pm$0.367 & 1.754$\pm$0.054 & 3.570$\pm$0.035 & 2.888$\pm$0.13 & 2.821$\pm$0.059\\ \hline
Rosalind & 6.290$\pm$0.129 & 5.136$\pm$0.336 & 3.823$\pm$0.086 & 0.477$\pm$0.017 & 1.031$\pm$0.029 & 0.935$\pm$0.101 & 0.716$\pm$0.039\\
(ring bg removed) & 6.208$\pm$0.131 & 5.038$\pm$0.352 & 3.843$\pm$0.083 & 0.472$\pm$0.021 & 1.001$\pm$0.026 & 0.918$\pm$0.110 & 0.673$\pm$0.044\\ \hline
Cupid & 0.400$\pm$0.038 & 0.287$\pm$0.009 & 0.250$\pm$0.020 & --- & ---& --- &--- \\ \hline
Belinda & 8.625$\pm$0.176 & 6.93$\pm$0.33 & 5.361$\pm$0.087 & 0.611$\pm$0.033 & 1.374$\pm$0.033 & 1.207$\pm$0.076 & 1.036$\pm$0.053\\ \hline
Perdita & 0.910$\pm$0.020 & 0.642$\pm$0.036 & 0.610$\pm$0.050 & --- & ---& --- & --- \\ \hline
Puck & 40.278$\pm$1.35 & 30.997$\pm$0.502 & 24.304$\pm$0.608 & 2.118$\pm$0.043 & 5.142$\pm$0.137 & 3.889$\pm$0.132 & 4.004$\pm$0.057\\ \hline
Mab & 0.696$\pm$0.028 & 0.459$\pm$0.017 & 0.281$\pm$0.020 & --- &--- & --- & --- \\ \hline
Miranda & 912.36$\pm$53.90 & 563.51$\pm$14.50 & 421.93$\pm$11.38 & 14.61$\pm$0.34 & 38.69$\pm$0.17 & 23.52$\pm$0.17 & 28.15$\pm$0.21\\
\hline\hline
Naiad & 1.253$\pm$0.023 & --- & 0.859$\pm$0.067 & 0.147$\pm$0.01 & --- & --- & --- \\ \hline
Thalassa & 2.021$\pm$0.286 & --- & 1.310$\pm$0.049 & 0.172$\pm$0.014 & --- & --- & --- \\ \hline
Despina & 6.962$\pm$0.217 & --- & 5.055$\pm$0.246 & 0.965$\pm$0.051 & --- & 1.25$\pm$0.137 & --- \\ \hline
Galatea & 8.324$\pm$0.546 & --- & 5.945$\pm$0.143 & 1.260$\pm$0.030 & --- & 1.395$\pm$0.140 & --- \\ \hline
Larissa & 10.160$\pm$0.282 & --- & 6.992$\pm$0.158 & 1.469$\pm$0.056 & --- & 1.672$\pm$0.276 & ---\\ \hline
Hippocamp & 0.288$\pm$0.056 & --- & 0.167$\pm$0.024 & --- & --- & --- & ---\\ \hline
Proteus & 57.282$\pm$1.755 & --- & 40.481$\pm$0.178 & 9.061$\pm$0.111 & --- & 9.406$\pm$0.158 & ---\\ \hline
\end{tabular}}
\end{table}

\begin{table}
\caption{Average  reflectances of the observed moons. Note all numbers are the measured values  at the observed phase angles (2.8$^\circ$-2.9$^\circ$ for Uranus, 1.8$^\circ$ for Neptune, see Table~\ref{geomtab}). {For the inner Uranian moons, the first row is without removal of the ring background and the second row is after removing the ring background.}}
\label{moontab2}
\hspace{-.5in}\resizebox{\textwidth}{!}{\begin{tabular}{|c|c|c|c|c|c|c|c|c|}
\hline
Moon & r (km) & F140M & F182M & F210M & F300M & F410M & F460M & F480M \\ 
\hline
Bianca & 28$^a$ & 0.0618$\pm$0.0014 & 0.0614$\pm$0.0011 & 0.0612$\pm$0.0022 & 0.0132$\pm$0.0028 & 0.0551$\pm$0.0052 & 0.0421$\pm$0.0072 & 0.0577$\pm$0.0090\\
(ring bg removed) & & 0.0621$\pm$0.0013 & 0.0623$\pm$0.0010 & 0.0609$\pm$0.0020 & 0.0180$\pm$0.0028 & 0.0559$\pm$0.0021 & 0.0507$\pm$0.0028 & 0.0684$\pm$0.0100\\ \hline
Cressida & 41$^a$ &  0.0661$\pm$0.0020 & 0.0656$\pm$0.0009 & 0.0628$\pm$0.0014 & 0.0148$\pm$0.0013 & 0.0468$\pm$0.0009 & 0.0576$\pm$0.0062 & 0.0517$\pm$0.0017\\
(ring bg removed) & & 0.0652$\pm$0.0020 & 0.0648$\pm$0.0010 & 0.0637$\pm$0.0013 & 0.0152$\pm$0.0014 & 0.0494$\pm$0.0014 & 0.0615$\pm$0.0074 & 0.0535$\pm$0.0023\\ \hline
Desdemona & 34$^a$ & 0.0771$\pm$0.0029 & 0.0764$\pm$0.0014 & 0.0729$\pm$0.0019 & 0.0164$\pm$0.0003 & 0.0568$\pm$0.0017 & 0.0615$\pm$0.0031 & 0.0622$\pm$0.0045\\
(ring bg removed) & & 0.0762$\pm$0.0029 & 0.0767$\pm$0.0013 & 0.0724$\pm$0.0026 & 0.0167$\pm$0.0006 & 0.0591$\pm$0.0019 & 0.0608$\pm$0.0058 & 0.0616$\pm$0.0053\\ \hline
Juliet & 53$^a$ & 0.0666$\pm$0.0035 & 0.0689$\pm$0.0006 & 0.0667$\pm$0.0016 & 0.0143$\pm$0.0008 & 0.0514$\pm$0.0007 & 0.0532$\pm$0.0016 & 0.0543$\pm$0.0019\\
(ring bg removed) &  & 0.0657$\pm$0.0035 & 0.0681$\pm$0.0005 & 0.0674$\pm$0.0016 & 0.0143$\pm$0.0007 & 0.0517$\pm$0.0008 & 0.0545$\pm$0.0014 & 0.0583$\pm$0.0025\\ \hline
Portia & 69$^a$ & 0.0712$\pm$0.0020 & 0.0697$\pm$0.0008 & 0.0667$\pm$0.0015 & 0.0154$\pm$0.0005 & 0.0557$\pm$0.0006 & 0.0569$\pm$0.0028 & 0.0581$\pm$0.0006\\
(ring bg removed) & & 0.0702$\pm$0.0019 & 0.0695$\pm$0.0009 & 0.0673$\pm$0.0015 & 0.0152$\pm$0.0005 & 0.0550$\pm$0.0005 & 0.0564$\pm$0.0025 & 0.0604$\pm$0.0013\\ \hline
Rosalind & 36$^a$ & 0.0696$\pm$0.0012 & 0.0722$\pm$0.0047 & 0.0646$\pm$0.0014 & 0.0154$\pm$0.0007 & 0.0583$\pm$0.0016 & 0.0671$\pm$0.0072 & 0.0563$\pm$0.0030\\
(ring bg removed) &  & 0.0687$\pm$0.0013 & 0.0709$\pm$0.005 & 0.0649$\pm$0.0013 & 0.0152$\pm$0.0008 & 0.0567$\pm$0.0015 & 0.0659$\pm$0.0079 & 0.0529$\pm$0.0035\\ \hline
Cupid & 9$^d$ & 0.0702$\pm$0.0064 & 0.0645$\pm$0.0021 & 0.0678$\pm$0.0054 & ---& --- & --- & --- \\\hline
Belinda & 43$^a$ & 0.0663$\pm$0.0013 & 0.0683$\pm$0.0033 & 0.0636$\pm$0.0010 & 0.0137$\pm$0.0008 & 0.0545$\pm$0.0013 & 0.0607$\pm$0.0038 & 0.0571$\pm$0.0029\\ \hline
Perdita & 15$^a$ & 0.0582$\pm$0.0015 & 0.052$\pm$0.0029 & 0.0593$\pm$0.0048 & --- & --- & --- & --- \\ \hline
Puck & 81$^a$ & 0.0869$\pm$0.0027 & 0.0861$\pm$0.0014 & 0.0809$\pm$0.0019 & 0.0134$\pm$0.0002 & 0.0575$\pm$0.0015 & 0.0551$\pm$0.0019 & 0.0622$\pm$0.0009\\ \hline
Mab & 12$^{d,e}$ & 0.0687$\pm$0.0022 & 0.0581$\pm$0.0021 & 0.0428$\pm$0.0030 & --- & --- &--- & --- \\ \hline
Miranda & 236$^c$ & 0.2308$\pm$0.0128 & 0.1848$\pm$0.0048 & 0.1662$\pm$0.0044 & 0.0109$\pm$0.0003 & 0.0510$\pm$0.0002 & 0.0393$\pm$0.0003 & 0.0516$\pm$0.0004\\
\hline \hline
Naiad & 33$^b$ &  0.0874$\pm$0.0016 &--- & 0.0922$\pm$0.0072 & 0.0298$\pm$0.0021 & --- & --- & --- \\ \hline
Thalassa & 40$^b$ &  0.0959$\pm$0.0136 &--- & 0.0957$\pm$0.0036 & 0.0238$\pm$0.0020 & --- & --- & --- \\ \hline
Despina & 74$^b$ &  0.0965$\pm$0.0030 & --- & 0.1079$\pm$0.0052 & 0.0390$\pm$0.0021 & --- & 0.1153$\pm$0.0126 & --- \\ \hline
Galatea & 79$^b$ &  0.1012$\pm$0.0066 & --- & 0.1113$\pm$0.0027 & 0.0447$\pm$0.0011 & --- & 0.1129$\pm$0.0113 & --- \\ \hline
Larissa & 96$^b$ &  0.0837$\pm$0.0023 & --- & 0.0886$\pm$0.0020 & 0.0353$\pm$0.0013 & --- & 0.0917$\pm$0.0151 & --- \\ \hline
Hippocamp & 17$^f$ &  0.0756$\pm$0.0148 & --- & 0.0676$\pm$0.0098 & --- & --- & --- & --- \\ \hline
Proteus & 208$^b$ &  0.1005$\pm$0.0031 & ---& 0.1093$\pm$0.0005 & 0.0463$\pm$0.0006 & --- & 0.1098$\pm$0.0018 & ---\\ \hline
\end{tabular}}

$^a$ From \citet{Karkoschka01b}

$^b$ From \citet{Karkoschka03}

$^c$ From \citet{Thomas88}

$^d$ From \citet{Showalter06}

$^e$ Note that if Mab is a more ice-rich body, as indicated by its spectrum, its radius could be closer to 6 km \citep{Molter23}, yielding much higher reflectances.

$^f$ From \citet{Showalter19}
\end{table}

Table~\ref{moontab1} provides the average fluxes for all the moons considered in this study. The uncertainties on these values correspond to the standard error on the mean of all the individual flux measurements from different images. The individual fit parameters and estimates of the moon fluxes and reflectances are provided in the Appendix to this paper. Note that for the sake of simplicity these averages do not include corrections for the slight variations in the distances to the planets (see Table~\ref{geomtab}). 

For the Uranian moons, it is worth noting that we provide two sets of flux values for Bianca, Cressida, Desdemona, Juliet, Portia and Rosalind: one value computed using the original image and another with the model of the ring background removed. The background subtraction has little effect on the flux values in the F140M, F182M and F210M filters, but does produce slightly increased fluxes at long wavelengths for Bianca and Cressida, which are the two moons closest to the main rings. Even so, the differences are  generally comparable to the uncertainties. We were also able to obtain flux measurements of  the small moons Cupid, Perdita and Mab from images obtained with the F140M, F182M and F210M filters. We did search for potential signals from all these moons at longer wavelengths, but were unable to find any convincing detections.  Finally,  we also include flux measurements for the innermost large moon Miranda, which is the only one of Uranus' five large moons that was not saturated in the shorter-wavelength filter images. 

Turning to Neptune's moons, we first note that these were only observed at four of the seven wavelengths. We could obtain flux measurements of the four larger moons (Proteus, Larissa, Galatea and Despina) at all four wavelengths. However, the smaller moons Naiad and Thalassa could not be clearly detected in the F460M images due to their limited signal-to-noise (note that both these moons were obscured by Neptune's disk during the longer-exposure BRIGHT1 images using this particular filter). We were also able to measure the flux of the smallest inner moon Hippocamp in the images obtained with the F140M and F210M filters. 

Again, since we are observing reflected light from the sun, it is useful to convert these flux values into measures of the moons' average reflectance $\bar{R}$. which is given by the following expression:

\begin{equation}
\bar{R}=\frac{\mathcal{F}_{moon}}{\mathcal F_\Sun}\left(\frac{D_{\sun}}{1 AU}\right)^2\frac{D_{obs}^2}{r^2}
\end{equation}
where $\mathcal{F}_{moon}$ is the measured moon flux, $\mathcal{F}_\sun$ is the appropriate solar flux at 1 AU provided in Table~\ref{filtertab}, $D_\sun$ and $D_{obs}$ are the distances from the moon to the Sun and the observer, respectively, which are provided in Table~\ref{geomtab}, and $r$ is the moon's effective average radius, which are included in Table~\ref{moontab2} {(note that the $D_{obs}$ values in Table~\ref{geomtab} are converted to kilometers for the purposes of calculating $\bar{R}$)}. For most of the moons we use radius estimates derived from resolved Voyager images \citep{Thomas88, Karkoschka01b, Karkoschka03}, while the sizes of Cupid, Mab and Hippocamp are based on estimates from HST observations assuming these moons have similar reflectances to their neighbors \citep{Showalter06, Showalter19}. This assumption is particularly problematic for Mab, which has a very different spectra from the other small moons, indicating that it has a more ice-rich spectra that would imply a radius closer to 6 km \citep{Molter23}. We compute the reflectances for each moon in every individual image (accounting for the slight variations in the observation geometry) and provide the average values at each relevant wavelength in Table~\ref{moontab2}. Again, the uncertainties on these estimates correspond to the standard error on the mean of the individual measurements. While this quantity is sometimes referred to as albedo, we prefer the term reflectance in order to avoid potential confusion with the geometric albedo, which includes an additional factor to correct the observed flux to its predicted values at a phase angle of 0$^\circ$ \citep{Emery07}. {Also note that our values for $\bar{R}$ are directly comparable to the albedo parameter computed by \citet{Belyakov24}.}

 \begin{figure}
\resizebox{0.9\textwidth}{!}{\includegraphics{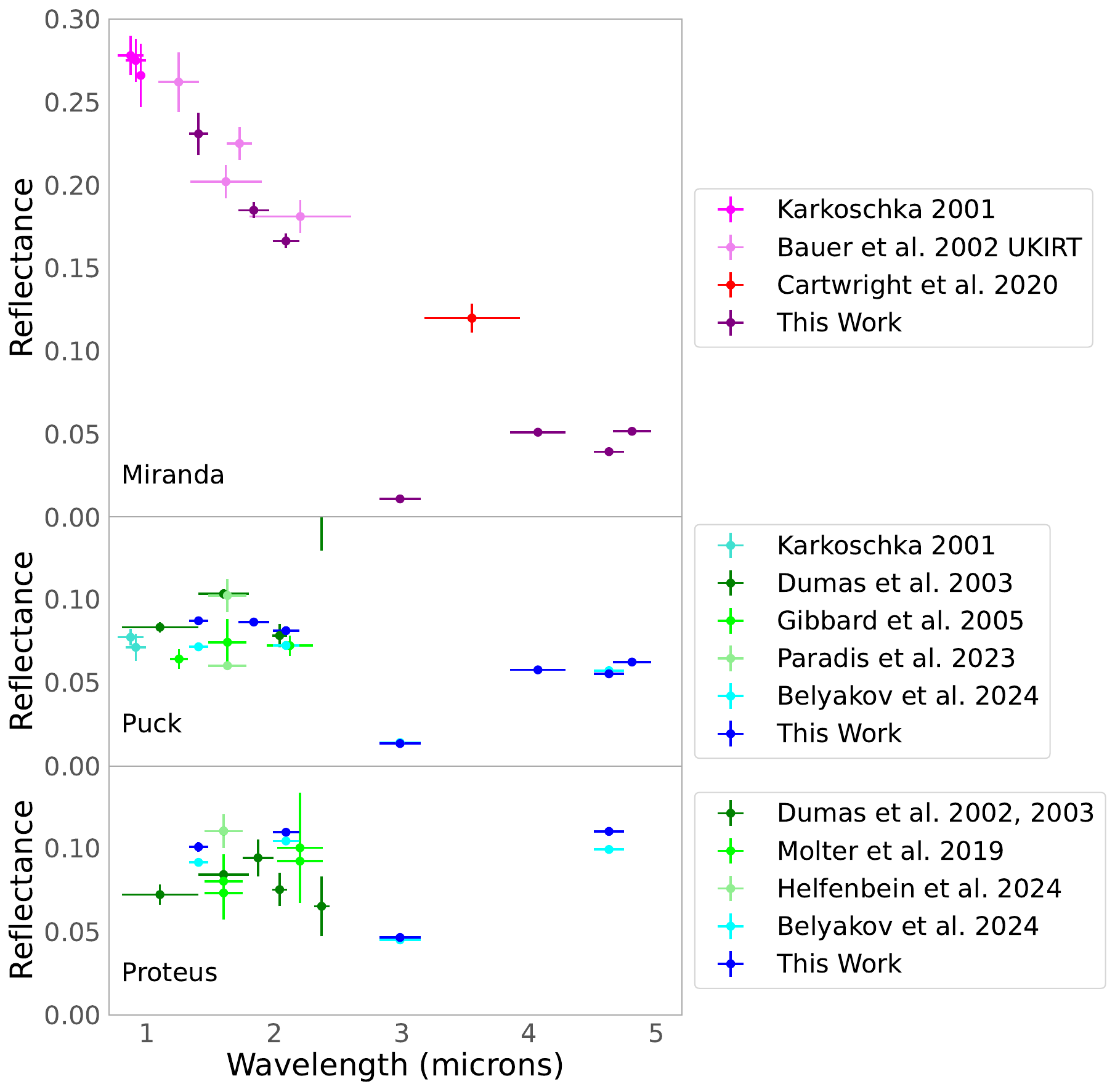}}
\caption{Comparisons of the new JWST estimates of the reflectances of Miranda, Puck and Proteus with previously published values. Note that in order to avoid complications associated with accounting for the moons' opposition surges, these plots show reflectance estimates derived from observations taken at phase angles comparable to those observed by JWST.  More specifically, for Uranus the plotted data from \citet{Karkoschka01}, \citet{Bauer02} and \citet{Cartwright20a} were all obtained at phase angles between 2.5$^\circ$ and 2.9$^\circ$. These values are therefore all close to the phase angles of 2.8-2.9$^\circ$ observed by JWST.  (Note that we only plot the UKIRT data from \citet{Bauer02} because these had the tightest error bars, and the published brightness estimate from \citet{Cartwright20a}  includes a phase correction that we removed in this plot). The Uranus data derived from \citet{Dumas03}, \citet{Gibbard05} and \citet{Paradis23} were obtained at phase angles between 1.6$^\circ$ and 2$^\circ$, which  are a bit lower than those observed by JWST. We chose not to correct these  data for phase angle variations using the \citet{Karkoschka01} formulae because there is not an obvious strong trend with phase angle among these observations.  Similarly, the data for Proteus from  \citet{Dumas02, Dumas03}, \citet{Molter19} and \citet{Helfenbein24}  were obtained at phase angles between 0.7$^\circ$ and 1.9$^\circ$, which are generally comparable to the phase angle of 1.8$^\circ$ observed by  JWST. Again, there is no obvious trend {over the limited observed range of phase angles} that would motivate a phase correction for these data.}
\label{speccomp}
\end{figure}
 
\section{Results}
\label{results}

For this particular study, we will not focus on the absolute brightnesses of the rings and moons because these parameters are more sensitive to the uncertain sizes and shapes of the various moons, as well as the observed phase angle and the accuracy of our correction factors. Instead we will simply show that our brightness estimates are reasonably consistent with previous published values  \citep{Karkoschka01, Karkoschka03, Dumas02,Dumas03,Gibbard05,Cartwright20a,Molter19,Molter23,Paradis23, Helfenbein24} and are not too far off the previous analyses of the JWST data by \citet{Belyakov24}.  Figure~\ref{speccomp} provides some  illustrative comparisons of our derived reflectance estimates for Miranda, Puck and Proteus with previously published values obtained from observations at comparable phase angles. First, note that our brightness estimates for Puck and Proteus are all within $\sim$10\% of the values derived by \citet{Belyakov24}.  The differences between these two sets of estimates are larger than their statistical error bars and probably arise from the differences in the data reduction techniques discussed above. Still, these differences are smaller than the dispersion of the earlier published measurements at wavelengths less than 2.5 $\mu$m \citep{Karkoschka01, Dumas02, Dumas03, Gibbard05, Molter19, Paradis23, Helfenbein24}, meaning that both estimates are reasonably compatible with prior measurements.  It is  also worth noting that our reflectance estimates for Miranda at wavelengths less than 2.5 $\mu$m are consistent with previously published measurements at comparable phase angles by \citet{Karkoschka01} and \citet{Bauer02}, providing further evidence that our brightness estimates are reasonable. Interestingly, we find that Miranda has a similar reflectance as the smaller moons like Puck at wavelengths longer than 2.5 $\mu$m. This result might at first  appear to be inconsistent with the much higher estimates of Miranda's brightness around 3.6 $\mu$m derived from Spitzer IRAC data by \citet{Cartwright20a}. However, the JWST and IRAC data can be consistent with each other if Miranda has a water-ice-rich spectrum similar to Ariel. \citet{Cartwright24} found that Ariel's spectrum shows a strong peak around 3.6 $\mu$m that is characteristic of water ice and would naturally cause the brightness around that wavelength to be over two times higher than it would be at the nearby wavelengths observed by JWST NIRCam.  This aspect of Miranda's spectrum therefore supports the idea that this moon has a water-ice-rich surface, which is relevant to the broader trends in the spectral properties of the Uranian rings and moons discussed in more detail below.

\begin{figure}
\resizebox{\textwidth}{!}{\includegraphics{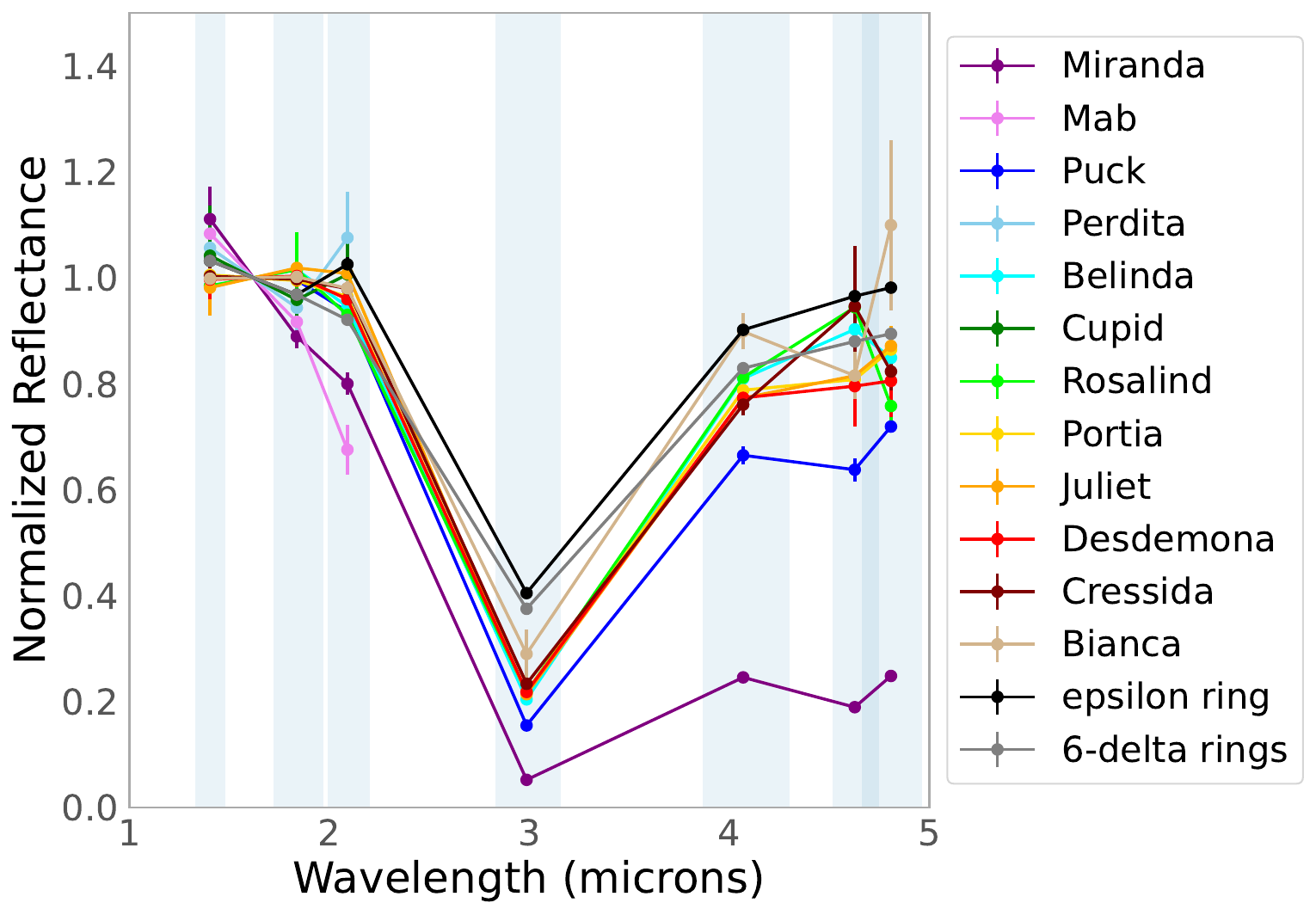}\includegraphics{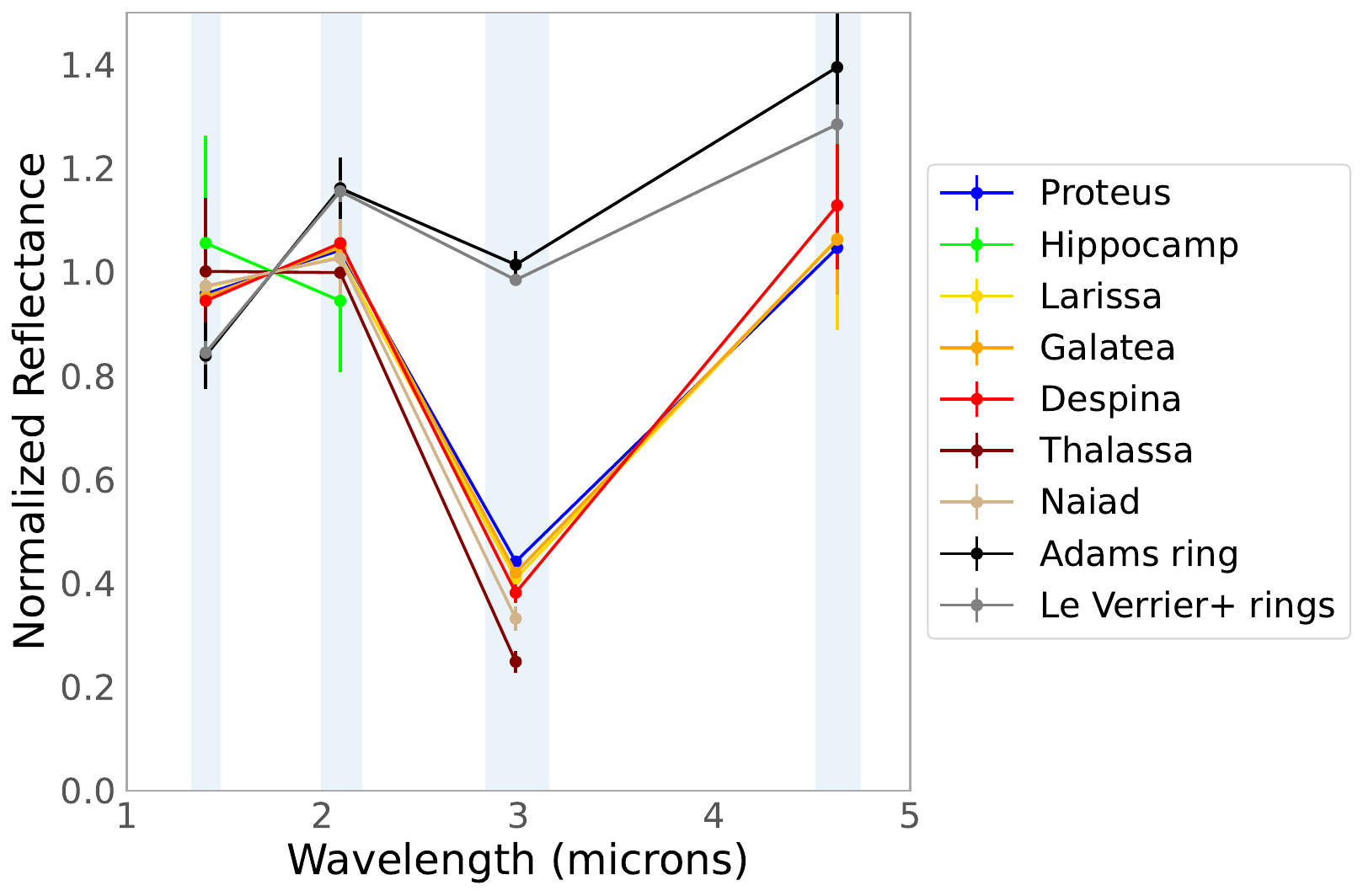}}
\caption{Normalized spectra of the rings and inner moons of Uranus and Neptune. All spectra are normalized to have an average value of 1 between 1.4 $\mu$m and 2.1 $\mu$m in order to facilitate comparisons among the different spectra. Note that while each data point corresponds to the average brightness over a range of wavelengths indicated by the vertical bands (see Table~\ref{filtertab}).}
\label{specs1}
\end{figure}

\begin{figure}
\resizebox{\textwidth}{!}{\includegraphics{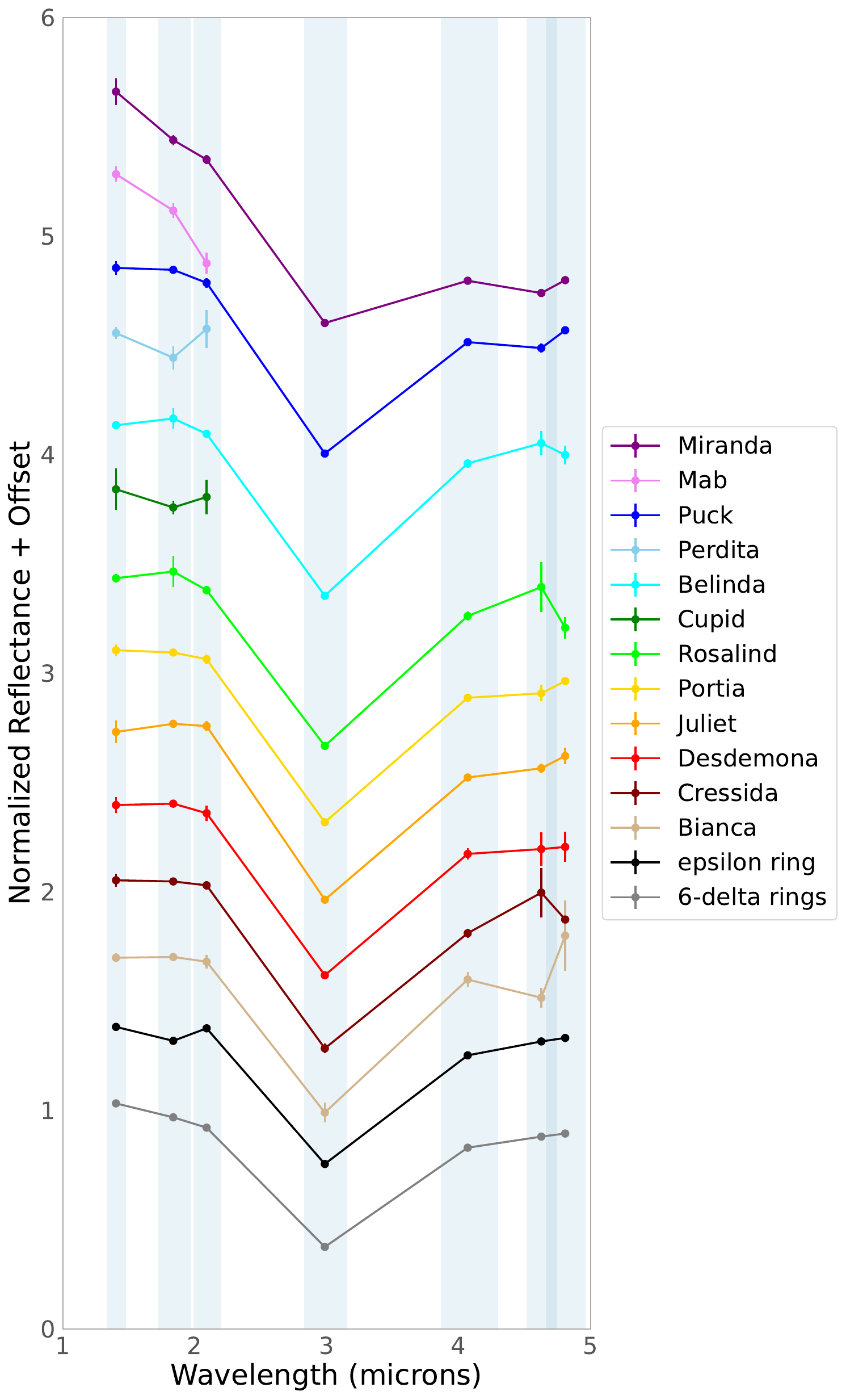}\includegraphics{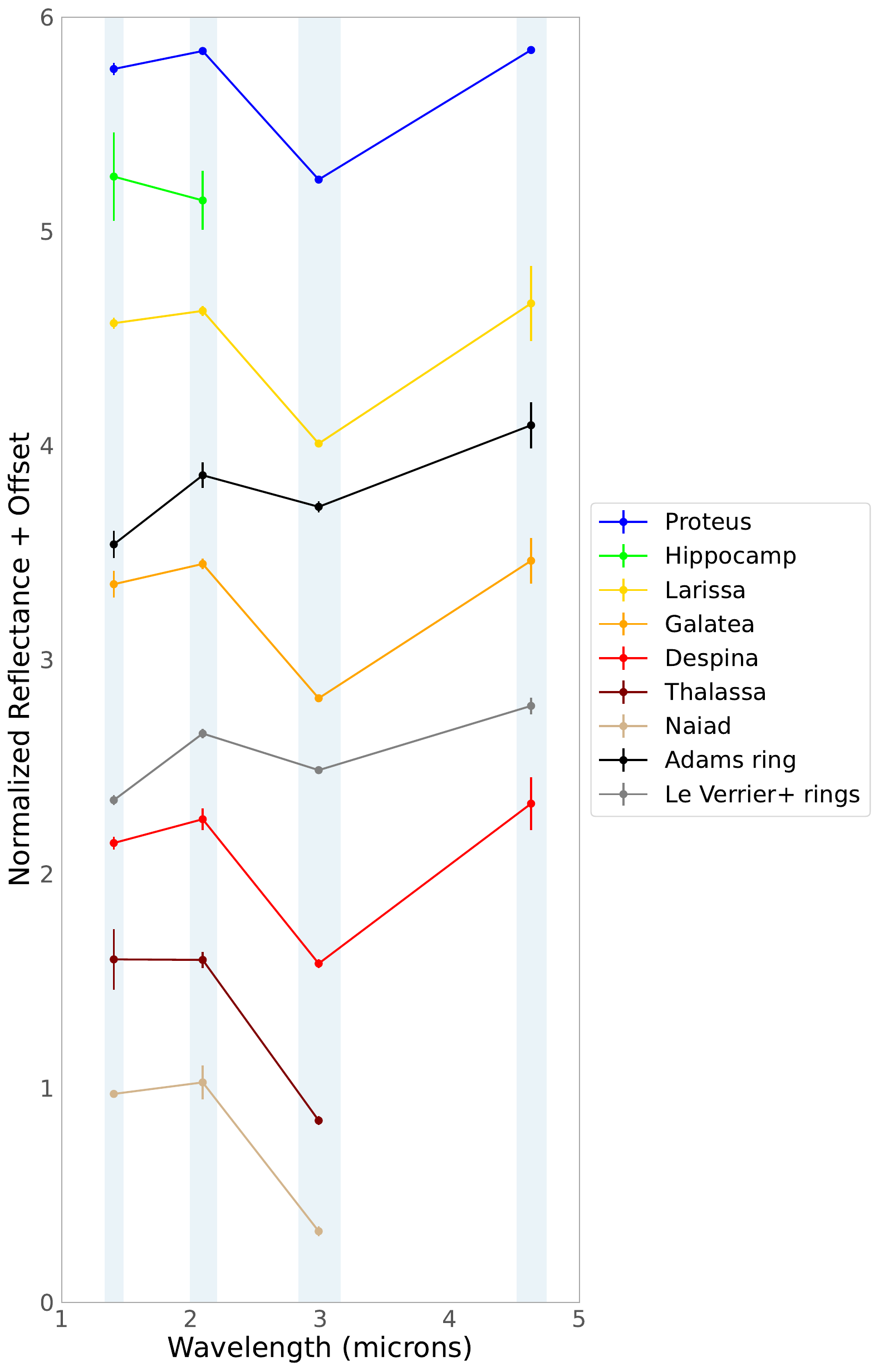}}
\caption{Normalized spectra of the rings and inner moons of Uranus and Neptune. These spectra are again all normalized to have an average value of 1 between 1.4 $\mu$m and 2.1 $\mu$m, but has also been offset vertically to illustrate trends with distance from the two planets.}
\label{specs2}
\end{figure}

\begin{figure}
\resizebox{\textwidth}{!}{\includegraphics{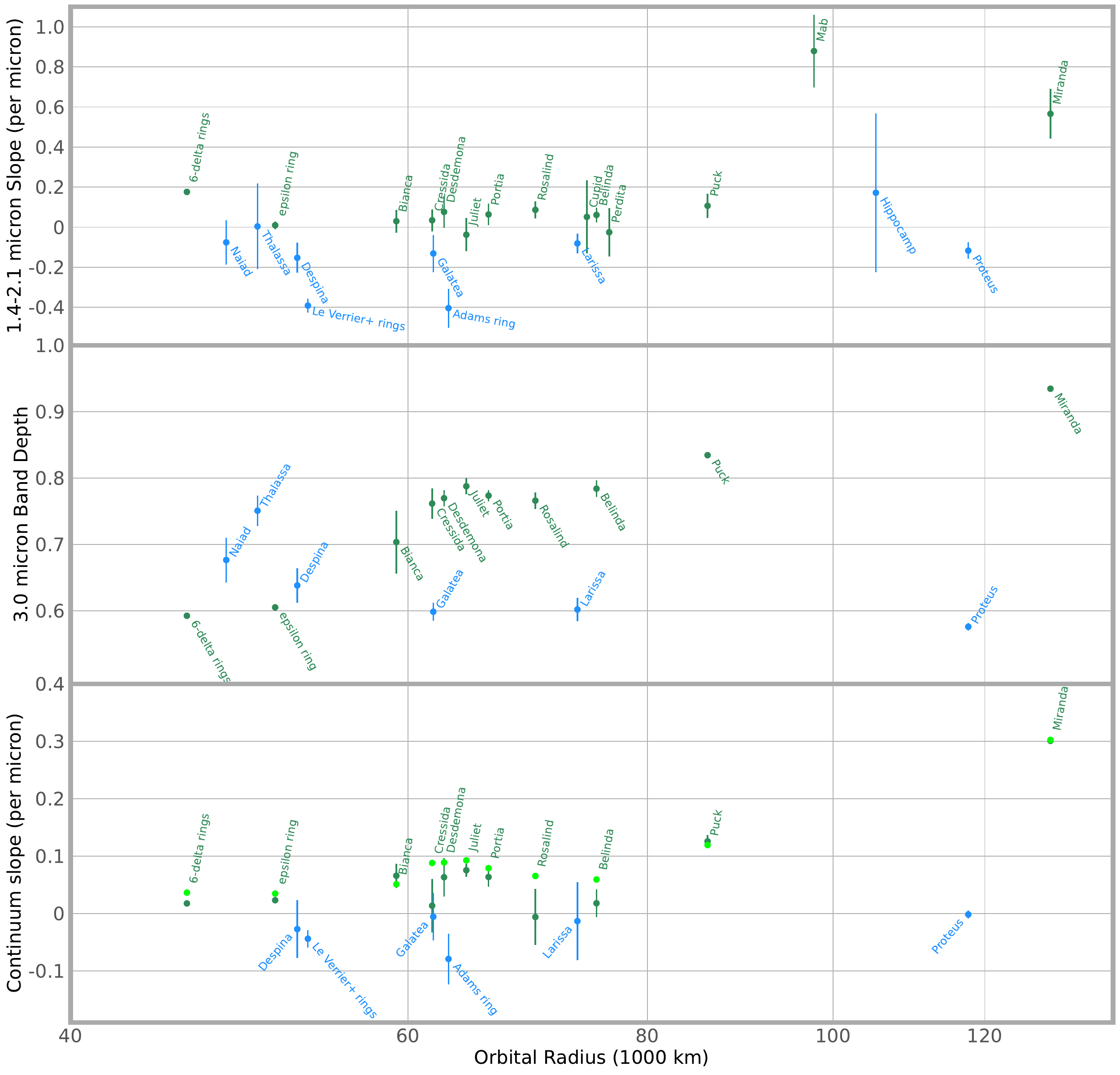}}
\caption{Plots of various spectral parameters of the rings and inner moons as functions of distance from their host planet. The rings and moons of Uranus are shown in green and the rings and moons of Neptune are shown in the blue. The top panel shows the spectral slope between 1.4 $\mu$m and 2.1 $\mu$m, which is defined such that blue slopes are positive and red slopes are negative. The middle panel shows the depth of the 3$\mu$m band, with larger values indicating a deeper band. Finally, the bottom panel shows the continuum slope across the two regions on either sides of the 3 $\mu$m band. The green and blue points just use the brightness data at 2.1 $\mu$m and 4.6 $\mu$m, while the lime points use the weighted average brightness of all three wavelengths in both regions.}
\label{parplot}
\end{figure}

The primary focus of this study will be normalized spectra and brightness ratios between various wavelengths.\footnote{Note that for this discussion we will also focus on the versions of the Uranus brightness measurements with the ring background removed} Figures~\ref{specs1} and~\ref{specs2} show the spectra for the various rings and moons, which have all been normalized so that the average signal between 1.4 $\mu$m and 2.1 $\mu$m is equal to 1, which makes the differences in the spectral shapes easier to compare. Note that for most of these spectra, the most prominent feature is a clear dip at 3 $\mu$m, which corresponds to the fundamental OH band that is common in many objects in the outer solar system \citep{dePra23, Emery23, Belyakov24}. In addition, there are variable slopes on either side of this band.

Figure~\ref{parplot} summarizes three spectral parameters that can be extracted from these normalized spectra. Each of these parameters is computed based on the ratios of the average reflectances provided in Tables~\ref{ringtab} and~\ref{moontab2}, which we will generically designate $\bar{R}_{FXXXM}$, where $FXXXM$ corresponds to the appropriate NIRCAM filter. The specific parameters discussed below were chosen because they provide information about the most obvious features in the spectra, and  because most of them could be computed for material orbiting both Uranus and Neptune, which facilitates comparisons between the two systems.

The top panel of Figure~\ref{parplot} shows the 1.4-2.1 $\mu$m spectral slope $S_{1.4/2.1}$, which is given by the following expression.

\begin{equation}
S_{1.4/2.1} ={\left(\frac{\bar{R}_{F140M}}{\bar{R}_{F210M}}-1\right)}\frac{1}{2.090 \mu m- 1.403 \mu m}
\end{equation}
Note that this spectral slope is defined so that positive values of $S_{1.4/2.1}$ correspond to blue spectral slopes over this portion of the spectrum, while negative values correspond to red spectral slopes.

The middle panel of Figure~\ref{parplot} shows the 3 $\mu$m band depth $BD_{3}$, which is given by the following expression

\begin{equation}
BD_{3} ={\left(1-\frac{\bar{R}_{F300M}}{\bar{R}_{F210M}}\right)}
\end{equation}
This quantity ranges between 0 and 1, with larger values corresponding to a deeper absorption around 3 $\mu$m.

The bottom panel of Figure~\ref{parplot} shows the broader continuum spectral slope $S_c$, which is given by the following expression.

\begin{equation}
S_c ={\left(\frac{\bar{R}_{F460M}}{\bar{R}_{F210M}}-1\right)}\frac{1}{2.090 \mu m- 4.627 \mu m}
\end{equation}
This parameter measures the overall spectral slope between the two regions on either side of the 3 $\mu$m band. Again,  this spectral slope is defined so that positive values of $S_c$ correspond to blue spectral slopes over this portion of the spectrum, while negative values correspond to red spectral slopes. For Uranus, we also include a plot of the related quantity:

\begin{equation}
S^*_c ={\left(\frac{\bar{R}_{long}}{\bar{R}_{short}}-1\right)}\frac{1}{\lambda_{short}- \lambda_{long}}
\end{equation}
where $\bar{R}_{short}$ is the weighted average of $\bar{R}_{F140M}$, $\bar{R}_{F182M}$ and $\bar{R}_{F210M}$,  $\bar{R}_{long}$ is the weighted average of $\bar{R}_{F410M}$, $\bar{R}_{F460M}$ and $\bar{R}_{F480M}$, and $\lambda_{short}$ and $\lambda_{long}$ are the corresponding weighted averages of the relevant filter wavelengths from Table~\ref{filtertab}. Due to the larger number of brightness measurements incorporated into $S^*_c$, this quantity has a smaller statistical uncertainty than $S_c$. 

{It is also worth noting that for the Uranian rings and moons, one could also define parameters that quantify the curvature of the spectra in the regions on either side of the 3-$\mu$m band. One of these parameters  would be the ratio of observed brightness in the F182M filter to the predicted brightness based on a a linear interpolation between the brightness measurements in the F140M and F210M filters. The other parameter would be the ratio of the observed signal in the F460M filter to the predicted signal based on  a linear interpolation between the F410M and F480M filters. Despite the observed spectra showing some potentially interesting variations in these parameters, we will not consider these parameters in detail here. One reason for this choice is that these parameters cannot be computed for Neptune's rings and moons. In addition, variations in these parameters are more difficult to securely interpret  both because the variations are relatively subtle and because the parameters themselves involve combining data obtained from different observations. We therefore leave analysis of those aspects of these spectra to future work.}

\section{Discussion of notable spectral trends}
\label{discussion}

{The spectral properties of rings and moons depend upon both their chemical compositions and the typical pathlength the observed light took through the relevant material. In principle,  one can obtain constraints on both these surface properties with sophisticated modeling of sufficient spectral data. Such detailed spectral modeling is beyond the scope of this work, and so we will instead focus on variations in the observed spectral properties and highlight interesting trends that merit further investigation. When appropriate, we do discuss whether the observed trends might be due to variations in the surface composition or the particle sizes of the rings and moons.}

The first thing to note about these spectra is that most of the small moons around each of the Ice Giants have very similarly shaped spectra. In particular, the normalized spectra of the Uranian moons Cressida, Desdemona, Juliet, Portia, Rosalind and Belinda are very close to each other. Similarly, the normalized spectra of Neptune's moons Despina, Galatea, Larissa and Proteus are very similar to each other. This degree of consistency across moons of a variety of sizes and distances from the planet indicates that our methods of isolating and quantifying the signals from the moons are robust, and so gives us more confidence that the variations among these spectra are physically meaningful.

The most obvious difference among the spectral properties of these objects is that Neptune's rings have a much weaker 3-$\mu$m band than either Uranus' rings or any of either planet's moons. The most likely explanation for this finding is that Neptune's rings are composed primarily of dust-sized ($<$ 100 $\mu$m) particles \citep{Porco95, dePater18}.  Such small particles only provide short pathlengths of material to the incident light, which naturally suppresses the absorption bands in reflected spectra \citep{Hedman18}. In addition, Neptune's rings have a redder 1.4-2.1 $\mu$m slope than Neptune's moons, but the broader continuum slope between 2.1 and 4.6 $\mu$m is similar to that of the nearby moons (see Figure~\ref{parplot}). Red spectral slopes are common in dusty rings, but the exact value of the slope depends on the particle size distribution and composition of the ring particles. Detailed modeling of these slopes should therefore provide constraints on the particle properties of these rings.

There are also clear systematic differences between the small moons of Neptune and Uranus that can be seen in Figure~\ref{parplot}. More specifically, the 3-$\mu$m absorption band is stronger on Uranus' moons than it is on most of Neptune's moons, and both the spectral slopes are bluer for the Uranian moons than they are for Neptune's moons. This is consistent with the previous analysis of the JWST NIRCam data by \citet{Belyakov24}, and suggests that there are systematic compositional differences in the material orbiting Uranus and Neptune, with the Uranian material probably having a higher fraction of water ice. 

The most novel findings from this new analysis are the trends in spectral parameters within each system. The clearest of these trends are in the 3-$\mu$m band depth shown in Figure~\ref{parplot}. Interestingly, the two systems show opposite trends, with the 3-$\mu$m band becoming stronger with increasing distance from the planet for the Uranian rings and moons, while for Neptune's moons this band appears to become weaker at larger distances. These trends could potentially indicate a convergence in the composition of the circumplanetary material close to the two planets. While detailed investigations of the implications of these broad spectral trends for the evolution of these systems is beyond the scope of this paper, we can highlight several details of these trends that merit further consideration as part of these efforts.

For one, the dense Uranian rings have 3-$\mu$m bands that are weaker than any of the Uranian small moons and are instead comparable to the bands of Neptune's outer small moons. While the very weak 3-$\mu$m band in Neptune's rings can be attributed to the small size of the particles in those rings, the relative weakness of the 3-$\mu$m band in the dense Uranian rings is unlikely to be purely a particle size effect because the dense Uranian rings are composed primarily of particles in the millimeter to meter size range \citep{Esposito91, French91, Nicholson18}.   The particles in these rings are therefore comparable in size to those found in Saturn's main rings, and Saturn's main rings have stronger water-ice absorption bands than most of Saturn's moons \citep{Filacchione13, Miller24, Ciarniello24}. Hence the difference in band depth between Uranus' rings and its moons could represent the endpoint of the broader compositional gradient in this system, implying that the rings close to the planet have a lower water-ice fraction than most of the small moons. This interpretation would also be consistent with the lower albedos of the rings compared to the inner moons \citep{Karkoschka01b}, and perhaps even with the higher density of the moons closest to the rings \citep{French24}. If this compositional trend between the rings and small moons can be confirmed, it could provide insights into how much the tidal evolution and disruption of material can influence the composition of circumplanetary material.

Meanwhile, not only is Miranda substantially brighter than the inner small moons at wavelengths shorter than 2.5 $\mu$m, it also has a much deeper 3-$\mu$m band and much bluer spectral slopes. {This is all consistent with the spectroscopic data showing that Miranda has a water-ice rich surface \citep{Bauer02}, and is also compatible with the spectral properties of other objects with surfaces rich in water-ice like Saturn's and Uranus' mid-sized moons or Haumea \citep{Emery05, Grundy06, Fraser09, PinillaAlonso09, Cartwright20a, Cartwright24}.} The only small inner moon that has spectral properties comparable to Miranda is the outermost small moon Mab, which shows a blue slope between 1.4 and 2.1 $\mu$m  similar to that seen on Miranda. This finding confirms  previous ground-based infrared observations of Mab \citep{Molter23}, and suggests that Mab has a surface that is more water-ice-rich than the small moons orbiting closer to the planet. The JWST data therefore confirm that the variations in the spectral properties of the Uranian moons cannot be entirely attributed to their sizes, and reinforce how important Mab is for understanding the apparent compositional gradient in the Uranian system. 

Furthermore, the spectrum of Puck reveals that it has a stronger 3-$\mu$m band than the other small inner moons, as well as a bluer continuum slope  due to it having a lower relative brightness at longer wavelengths (see Figure~\ref{specs1}). This suggests that Puck has more water-ice on its surface than the moons orbiting interior to it. While this could potentially be further evidence for the overall compositional gradient, it could also be due to material flowing between Mab and Puck via the $\mu$ ring. This dusty ring with an unusual blue color \citep{dePater06} appears to consist of debris generated by Mab \citep{Showalter06}, and the shape of its brightness profile indicates that its material is preferentially transported inwards towards Puck \citep{Showalter06, Hedman18}.  Material from the $\mu$ ring would probably preferentially strike the leading and anti-Uranian sides of Puck, so future studies of spectral trends across Puck itself should be able to clarify the importance of material transport for these moons.

Finally, there are some more subtle variations in the continuum slope of the Uranian rings in Figure~\ref{parplot} that could provide additional insights into how material might have been transported across this system. In general, the continuum slope becomes bluer between the rings and Miranda, consistent with the trend in the depth of the 3-$\mu$m and water-ice abundance discussed above. However, this trend appears to be interrupted in the vicinity of the moon Rosalind. Rosalind, Belinda, and perhaps Portia have redder continuum slopes than the moons orbiting immediately interior to them. Rosalind also appears to have a slightly weaker 3-$\mu$m band than Juliet, despite orbiting further from the planet. These spectral anomalies could potentially be due to the dusty $\nu$ ring, which has a red color and lies between the orbits of Portia and Rosalind  \citep{Showalter06}. This ring even appears to  overlap with Rosalind's orbit \citep{Showalter06}, and so material from this ring could be covering that moon, giving it a redder continuum slope and a weaker 3-$\mu$m band. However, it is not yet clear whether $\nu$-ring material could also affect the spectral properties of other moons like Belinda. Again, debris from the $\nu$ ring will probably not have a uniform distribution over the surfaces of moons like Rosalind, so future analyses should be able to clarify how much dusty ring material could be influencing the surface composition of these moons.

\section{Summary and Conclusions}
\label{Summary}

The main findings of this investigation of the spectral properties of the Ice Giants' rings and small moons are as follows:
\begin{itemize}
\item The JWST NIRCam spectra provide consistent and robust broad-band spectra of the rings and small moons orbiting around both Uranus and Neptune.
\item The rings and small moons of both Uranus and Neptune have detectable 3-$\mu$m OH absorption bands.
\item The 3-$\mu$m band is relatively weak for Neptune's rings, most likely because these rings are composed of dust-sized ($<100 \mu$m wide) particles.
\item  Uranus' small moons have a deeper 3-$\mu$m band than most of Neptune's small moons, which suggests that the objects in the inner Uranian system have a higher fraction of water ice than the objects in the inner Neptunian system. 
\item For Neptune's moons, the depth of the 3-$\mu$m band decreases with distance from the planet, so the innermost moons may have surfaces that are more water-ice rich.
\item In the Uranian system, the depth of the 3-$\mu$m band increases with distance from the planet from the main rings through the small moons out to Miranda. This implies that the water-ice fraction of these objects' surfaces increases with distance from the planet.
\item Uranus' outermost small moon Mab has {a spectral slope at wavelengths less than 2.5 $\mu$m that is as blue as Miranda's}, indicating that it has a more water-ice-rich surface than the other small moons. 
\item Uranus' moon Puck has a deeper 3-$\mu$m band and bluer continuum spectral slope than the moons orbiting interior to it. This could potentially be due to $\mu$-ring material originating from Mab covering the surface of this moon.
\item Uranus' moon Rosalind has a slightly redder continuum spectral slope and a slightly weaker 3-$\mu$m band than nearby moons orbiting closer to the planet, perhaps due to material from the spectrally red $\nu$ ring coating this moon.
\end{itemize}

These findings demonstrate that the spectral properties of the rings and small moons around Uranus and Neptune merit further investigation. For example, future studies of the spectra of moons like Mab, Puck and Rosalind should constrain the importance of ongoing material transport for the surface composition of these objects. Meanwhile, more detailed spectra of multiple objects within each system should reveal whether the variations in the 3-$\mu$m band depths are correlated with other spectral signatures, which would clarify how the surface compositions of these bodies vary with distance from the planet. These sorts of studies would provide important new information about  the evolution and perhaps even the origins of the circumplanetary material around the Ice Giants.

\acknowledgements{The authors thank all the people working on JWST NIRCam who enabled these data to be collected, which can be found at {\tt http://dx.doi.org/10.17909/qtzb-3439}. LNF was supported by STFC Consolidated Grant reference ST/W00089X/1.﻿ The authors would also like to thank the reviewers for their helpful comments.}

\clearpage

\section*{Appendix: Data Tables for Individual Moon Brightness Estimates}

\clearpage

\begin{table}
\caption{Bianca brightness data (after ring subtraction)}
\resizebox{6.5in}{!}{
\begin{tabular}{|c|c|c|c|c|c|c|c|}\hline
Filename & Filter & Peak & Width1 & Width2 & Tilt & Flux & Reflectance \\
 &  & (MJy/sr) & (pixel) & (pixel) & (radian) & ($\mu$Jy) &  \\ \hline
jw02739003001\_02101\_00001\_nrcb1\_i2d.fits & F140M & 18.348 & 0.931 & 0.879 & 1.097 & 3.35 & 0.0624 \\
jw02739003001\_02101\_00002\_nrcb1\_i2d.fits & F140M & 19.037 & 0.957 & 0.792 & 0.293 & 3.22 & 0.0600 \\
jw02739003001\_02101\_00003\_nrcb1\_i2d.fits & F140M & 15.682 & 0.893 & 1.054 & 1.571 & 3.30 & 0.0614 \\
jw02739011001\_02101\_00001\_nrcb1\_i2d.fits & F140M & 20.100 & 0.978 & 0.834 & 1.430 & 3.66 & 0.0653 \\
jw02739011001\_02101\_00002\_nrcb1\_i2d.fits & F140M & 18.012 & 0.848 & 0.989 & 0.000 & 3.37 & 0.0602 \\
jw02739011001\_02101\_00003\_nrcb1\_i2d.fits & F140M & 15.929 & 1.059 & 0.863 & 1.357 & 3.25 & 0.0580 \\
\hline
\end{tabular}}
\end{table}

\begin{table}
\caption{Bianca brightness data (without ring subtraction)}
\resizebox{6.5in}{!}{
\begin{tabular}{|c|c|c|c|c|c|c|c|}\hline
Filename & Filter & Peak & Width1 & Width2 & Tilt & Flux & Reflectance \\
 &  & (MJy/sr) & (pixel) & (pixel) & (radian) & ($\mu$Jy) &  \\ \hline
jw02739003001\_02101\_00001\_nrcb1\_i2d.fits & F140M & 18.118 & 0.875 & 0.917 & 0.000 & 3.29 & 0.0614 \\
jw02739003001\_02101\_00002\_nrcb1\_i2d.fits & F140M & 18.991 & 0.928 & 0.790 & 0.300 & 3.16 & 0.0588 \\
jw02739003001\_02101\_00003\_nrcb1\_i2d.fits & F140M & 15.533 & 0.891 & 1.030 & 1.571 & 3.23 & 0.0602 \\
jw02739011001\_02101\_00001\_nrcb1\_i2d.fits & F140M & 20.008 & 0.967 & 0.839 & 1.477 & 3.68 & 0.0656 \\
jw02739011001\_02101\_00002\_nrcb1\_i2d.fits & F140M & 17.892 & 0.850 & 0.983 & 0.000 & 3.39 & 0.0604 \\
jw02739011001\_02101\_00003\_nrcb1\_i2d.fits & F140M & 15.714 & 1.053 & 0.872 & 1.361 & 3.27 & 0.0583 \\
\hline
\end{tabular}}
\end{table}

\begin{table}
\caption{Cressida brightness data (after ring subtraction)}
\resizebox{6.5in}{!}{
\begin{tabular}{|c|c|c|c|c|c|c|c|}\hline
Filename & Filter & Peak & Width1 & Width2 & Tilt & Flux & Reflectance \\
 &  & (MJy/sr) & (pixel) & (pixel) & (radian) & ($\mu$Jy) &  \\ \hline
jw02739003001\_02101\_00001\_nrcb1\_i2d.fits & F140M & 38.225 & 1.062 & 0.919 & 0.000 & 8.33 & 0.0724 \\
jw02739003001\_02101\_00002\_nrcb1\_i2d.fits & F140M & 39.907 & 0.824 & 0.988 & 0.376 & 7.26 & 0.0630 \\
jw02739003001\_02101\_00003\_nrcb1\_i2d.fits & F140M & 38.838 & 0.823 & 1.007 & 1.252 & 7.19 & 0.0624 \\
jw02739003001\_02101\_00004\_nrcb1\_i2d.fits & F140M & 47.954 & 0.835 & 0.792 & 0.000 & 7.09 & 0.0616 \\
jw02739011001\_02101\_00002\_nrcb1\_i2d.fits & F140M & 45.017 & 0.830 & 0.972 & 0.344 & 8.11 & 0.0674 \\
jw02739011001\_02101\_00003\_nrcb1\_i2d.fits & F140M & 35.439 & 1.000 & 0.843 & 0.000 & 6.67 & 0.0555 \\
\hline
\end{tabular}}
\end{table}
\begin{table}
\caption{Cressida brightness data (without ring subtraction)}
\resizebox{6.5in}{!}{
\begin{tabular}{|c|c|c|c|c|c|c|c|}\hline
Filename & Filter & Peak & Width1 & Width2 & Tilt & Flux & Reflectance \\
 &  & (MJy/sr) & (pixel) & (pixel) & (radian) & ($\mu$Jy) &  \\ \hline
jw02739003001\_02101\_00001\_nrcb1\_i2d.fits & F140M & 38.268 & 1.060 & 0.917 & 0.000 & 8.43 & 0.0732 \\
jw02739003001\_02101\_00002\_nrcb1\_i2d.fits & F140M & 40.013 & 0.822 & 0.986 & 0.368 & 7.35 & 0.0638 \\
jw02739003001\_02101\_00003\_nrcb1\_i2d.fits & F140M & 38.896 & 0.823 & 1.007 & 1.251 & 7.30 & 0.0634 \\
jw02739003001\_02101\_00004\_nrcb1\_i2d.fits & F140M & 47.941 & 0.836 & 0.793 & 0.000 & 7.20 & 0.0626 \\
jw02739011001\_02101\_00002\_nrcb1\_i2d.fits & F140M & 45.129 & 0.830 & 0.972 & 0.354 & 8.25 & 0.0686 \\
jw02739011001\_02101\_00003\_nrcb1\_i2d.fits & F140M & 35.512 & 1.000 & 0.842 & 0.000 & 6.78 & 0.0564 \\
\hline
\end{tabular}}
\end{table}

\begin{table}
\caption{Desdemona brightness data (after ring subtraction)}
\resizebox{6.5in}{!}{
\begin{tabular}{|c|c|c|c|c|c|c|c|}\hline
Filename & Filter & Peak & Width1 & Width2 & Tilt & Flux & Reflectance \\
 &  & (MJy/sr) & (pixel) & (pixel) & (radian) & ($\mu$Jy) &  \\ \hline
jw02739011001\_02101\_00002\_nrcb1\_i2d.fits & F140M & 40.487 & 0.863 & 0.823 & 0.000 & 6.42 & 0.0777 \\
jw02739011001\_02101\_00003\_nrcb1\_i2d.fits & F140M & 27.438 & 1.012 & 0.917 & 0.000 & 5.69 & 0.0688 \\
jw02739011001\_02101\_00004\_nrcb1\_i2d.fits & F140M & 38.670 & 0.859 & 0.919 & 1.306 & 6.83 & 0.0826 \\
jw02739011001\_02101\_00005\_nrcb1\_i2d.fits & F140M & 39.339 & 1.002 & 0.711 & 0.000 & 6.26 & 0.0757 \\
jw02768003001\_02103\_00002\_nrcb4\_i2d.fits & F182M & 18.669 & 1.169 & 0.953 & 0.542 & 4.73 & 0.0747 \\
jw02768003001\_02103\_00001\_nrcb4\_i2d.fits & F182M & 18.863 & 1.207 & 0.982 & 0.510 & 5.09 & 0.0803 \\
\hline
\end{tabular}}
\end{table}
\begin{table}
\caption{Desdemona brightness data (without ring subtraction)}
\resizebox{6.5in}{!}{
\begin{tabular}{|c|c|c|c|c|c|c|c|}\hline
Filename & Filter & Peak & Width1 & Width2 & Tilt & Flux & Reflectance \\
 &  & (MJy/sr) & (pixel) & (pixel) & (radian) & ($\mu$Jy) &  \\ \hline
jw02739011001\_02101\_00002\_nrcb1\_i2d.fits & F140M & 40.461 & 0.862 & 0.823 & 0.000 & 6.50 & 0.0787 \\
jw02739011001\_02101\_00003\_nrcb1\_i2d.fits & F140M & 27.422 & 1.010 & 0.918 & 0.000 & 5.76 & 0.0697 \\
jw02739011001\_02101\_00004\_nrcb1\_i2d.fits & F140M & 38.625 & 0.859 & 0.919 & 1.299 & 6.91 & 0.0835 \\
jw02739011001\_02101\_00005\_nrcb1\_i2d.fits & F140M & 39.330 & 1.000 & 0.711 & 0.000 & 6.34 & 0.0767 \\
jw02768003001\_02103\_00002\_nrcb4\_i2d.fits & F182M & 18.567 & 1.147 & 0.939 & 0.544 & 4.58 & 0.0724 \\
jw02768003001\_02103\_00001\_nrcb4\_i2d.fits & F182M & 18.780 & 1.186 & 0.974 & 0.514 & 4.97 & 0.0785 \\
\hline
\end{tabular}}
\end{table}

\begin{table}
\caption{Juliet brightness data (after ring subtraction)}
\resizebox{6.5in}{!}{
\begin{tabular}{|c|c|c|c|c|c|c|c|}\hline
Filename & Filter & Peak & Width1 & Width2 & Tilt & Flux & Reflectance \\
 &  & (MJy/sr) & (pixel) & (pixel) & (radian) & ($\mu$Jy) &  \\ \hline
jw02739011001\_02101\_00001\_nrcb1\_i2d.fits & F140M & 65.109 & 1.049 & 0.944 & 0.028 & 14.39 & 0.0716 \\
jw02739011001\_02101\_00002\_nrcb1\_i2d.fits & F140M & 82.695 & 0.895 & 0.817 & 0.000 & 13.51 & 0.0672 \\
jw02739011001\_02101\_00003\_nrcb1\_i2d.fits & F140M & 57.263 & 1.076 & 0.811 & 0.000 & 11.16 & 0.0556 \\
jw02739011001\_02101\_00004\_nrcb1\_i2d.fits & F140M & 73.177 & 0.936 & 0.895 & 0.000 & 13.70 & 0.0682 \\
jw02768002001\_02103\_00004\_nrcb4\_i2d.fits & F182M & 47.243 & 0.892 & 1.089 & 0.291 & 10.43 & 0.0677 \\
jw02768002001\_02103\_00003\_nrcb4\_i2d.fits & F182M & 48.117 & 0.948 & 1.036 & 0.778 & 10.75 & 0.0697 \\
\hline
\end{tabular}}
\end{table}
\begin{table}
\caption{Juliet brightness data (without ring subtraction)}
\resizebox{6.5in}{!}{
\begin{tabular}{|c|c|c|c|c|c|c|c|}\hline
Filename & Filter & Peak & Width1 & Width2 & Tilt & Flux & Reflectance \\
 &  & (MJy/sr) & (pixel) & (pixel) & (radian) & ($\mu$Jy) &  \\ \hline
jw02739011001\_02101\_00001\_nrcb1\_i2d.fits & F140M & 65.154 & 1.049 & 0.943 & 0.027 & 14.61 & 0.0727 \\
jw02739011001\_02101\_00002\_nrcb1\_i2d.fits & F140M & 82.779 & 0.895 & 0.816 & 0.000 & 13.71 & 0.0683 \\
jw02739011001\_02101\_00003\_nrcb1\_i2d.fits & F140M & 57.235 & 1.077 & 0.811 & 0.000 & 11.33 & 0.0564 \\
jw02739011001\_02101\_00004\_nrcb1\_i2d.fits & F140M & 73.173 & 0.936 & 0.895 & 0.000 & 13.90 & 0.0692 \\
jw02768002001\_02103\_00004\_nrcb4\_i2d.fits & F182M & 47.155 & 0.890 & 1.089 & 0.291 & 10.48 & 0.0680 \\
jw02768002001\_02103\_00003\_nrcb4\_i2d.fits & F182M & 48.106 & 0.945 & 1.033 & 0.780 & 10.77 & 0.0699 \\
\hline
\end{tabular}}
\end{table}

\begin{table}
\caption{Portia brightness data (after ring subtraction)}
\resizebox{6.5in}{!}{
\begin{tabular}{|c|c|c|c|c|c|c|c|}\hline
Filename & Filter & Peak & Width1 & Width2 & Tilt & Flux & Reflectance \\
 &  & (MJy/sr) & (pixel) & (pixel) & (radian) & ($\mu$Jy) &  \\ \hline
jw02739003001\_02101\_00001\_nrcb1\_i2d.fits & F140M & 126.345 & 0.943 & 0.909 & 0.832 & 24.18 & 0.0742 \\
jw02739011001\_02101\_00003\_nrcb1\_i2d.fits & F140M & 120.012 & 0.925 & 0.792 & 0.779 & 19.64 & 0.0577 \\
jw02739011001\_02101\_00002\_nrcb1\_i2d.fits & F140M & 121.629 & 1.024 & 0.903 & 0.747 & 25.14 & 0.0738 \\
jw02739011001\_02101\_00001\_nrcb1\_i2d.fits & F140M & 144.852 & 0.939 & 0.771 & 0.868 & 23.45 & 0.0689 \\
jw02739011001\_02101\_00004\_nrcb1\_i2d.fits & F140M & 131.368 & 0.993 & 0.830 & 1.063 & 24.19 & 0.0710 \\
jw02739003001\_02101\_00004\_nrcb1\_i2d.fits & F140M & 160.376 & 0.747 & 0.844 & 0.000 & 22.57 & 0.0692 \\
\hline
\end{tabular}}
\end{table}
\begin{table}
\caption{Portia brightness data (without ring subtraction)}
\resizebox{6.5in}{!}{
\begin{tabular}{|c|c|c|c|c|c|c|c|}\hline
Filename & Filter & Peak & Width1 & Width2 & Tilt & Flux & Reflectance \\
 &  & (MJy/sr) & (pixel) & (pixel) & (radian) & ($\mu$Jy) &  \\ \hline
jw02739003001\_02101\_00001\_nrcb1\_i2d.fits & F140M & 126.278 & 0.941 & 0.910 & 0.786 & 24.50 & 0.0751 \\
jw02739011001\_02101\_00003\_nrcb1\_i2d.fits & F140M & 120.003 & 0.925 & 0.792 & 0.780 & 19.93 & 0.0585 \\
jw02739011001\_02101\_00002\_nrcb1\_i2d.fits & F140M & 121.641 & 1.024 & 0.903 & 0.747 & 25.50 & 0.0749 \\
jw02739011001\_02101\_00001\_nrcb1\_i2d.fits & F140M & 144.830 & 0.940 & 0.772 & 0.869 & 23.80 & 0.0699 \\
jw02739011001\_02101\_00004\_nrcb1\_i2d.fits & F140M & 131.395 & 0.993 & 0.830 & 1.062 & 24.54 & 0.0721 \\
jw02739003001\_02101\_00004\_nrcb1\_i2d.fits & F140M & 160.291 & 0.747 & 0.842 & 0.000 & 22.84 & 0.0700 \\
\hline
\end{tabular}}
\end{table}

\begin{table}
\caption{Rosalind brightness data (after ring subtraction)}
\resizebox{6.5in}{!}{
\begin{tabular}{|c|c|c|c|c|c|c|c|}\hline
Filename & Filter & Peak & Width1 & Width2 & Tilt & Flux & Reflectance \\
 &  & (MJy/sr) & (pixel) & (pixel) & (radian) & ($\mu$Jy) &  \\ \hline
jw02739003001\_02101\_00002\_nrcb1\_i2d.fits & F140M & 33.925 & 0.898 & 0.886 & 0.000 & 6.03 & 0.0679 \\
jw02739003001\_02101\_00003\_nrcb1\_i2d.fits & F140M & 29.539 & 1.067 & 0.915 & 0.000 & 6.44 & 0.0726 \\
jw02739003001\_02101\_00004\_nrcb1\_i2d.fits & F140M & 37.977 & 0.793 & 0.870 & 0.763 & 5.85 & 0.0659 \\
jw02739011001\_02101\_00004\_nrcb1\_i2d.fits & F140M & 36.480 & 0.941 & 0.802 & 0.602 & 6.15 & 0.0664 \\
jw02739011001\_02101\_00005\_nrcb1\_i2d.fits & F140M & 36.644 & 0.952 & 0.842 & 1.068 & 6.56 & 0.0708 \\
jw02768001001\_02103\_00003\_nrcb4\_i2d.fits & F182M & 20.151 & 1.027 & 0.976 & 0.000 & 4.60 & 0.0646 \\
\hline
\end{tabular}}
\end{table}
\begin{table}
\caption{Rosalind brightness data (without ring subtraction)}
\resizebox{6.5in}{!}{
\begin{tabular}{|c|c|c|c|c|c|c|c|}\hline
Filename & Filter & Peak & Width1 & Width2 & Tilt & Flux & Reflectance \\
 &  & (MJy/sr) & (pixel) & (pixel) & (radian) & ($\mu$Jy) &  \\ \hline
jw02739003001\_02101\_00002\_nrcb1\_i2d.fits & F140M & 33.909 & 0.901 & 0.889 & 0.000 & 6.15 & 0.0693 \\
jw02739003001\_02101\_00003\_nrcb1\_i2d.fits & F140M & 29.389 & 1.070 & 0.911 & 0.000 & 6.49 & 0.0731 \\
jw02739003001\_02101\_00004\_nrcb1\_i2d.fits & F140M & 37.916 & 0.831 & 0.829 & 0.000 & 5.92 & 0.0667 \\
jw02739011001\_02101\_00004\_nrcb1\_i2d.fits & F140M & 36.486 & 0.940 & 0.802 & 0.606 & 6.24 & 0.0673 \\
jw02739011001\_02101\_00005\_nrcb1\_i2d.fits & F140M & 36.686 & 0.952 & 0.841 & 1.063 & 6.65 & 0.0718 \\
jw02768001001\_02103\_00003\_nrcb4\_i2d.fits & F182M & 20.147 & 1.025 & 0.973 & 0.000 & 4.61 & 0.0647 \\
\hline
\end{tabular}}
\end{table}

\begin{table}
\caption{Cupid brightness data (without ring subtraction)}
\resizebox{6.5in}{!}{
\begin{tabular}{|c|c|c|c|c|c|c|c|}\hline
Filename & Filter & Peak & Width1 & Width2 & Tilt & Flux & Reflectance \\
 &  & (MJy/sr) & (pixel) & (pixel) & (radian) & ($\mu$Jy) &  \\ \hline
jw02739003001\_02101\_00001\_nrcb1\_i2d.fits & F140M & 3.304 & 0.787 & 0.500 & 0.000 & 0.29 & 0.0531 \\
jw02739003001\_02101\_00002\_nrcb1\_i2d.fits & F140M & 1.605 & 1.022 & 1.138 & 0.000 & 0.42 & 0.0762 \\
jw02739003001\_02101\_00003\_nrcb1\_i2d.fits & F140M & 2.063 & 0.767 & 0.855 & 0.282 & 0.31 & 0.0552 \\
jw02739003001\_02101\_00004\_nrcb1\_i2d.fits & F140M & 2.595 & 1.045 & 0.500 & 0.235 & 0.31 & 0.0554 \\
jw02739011001\_02101\_00001\_nrcb1\_i2d.fits & F140M & 2.555 & 0.958 & 0.548 & 0.000 & 0.30 & 0.0525 \\
jw02739011001\_02101\_00002\_nrcb1\_i2d.fits & F140M & 2.448 & 1.265 & 0.835 & 0.000 & 0.59 & 0.1013 \\
\hline
\end{tabular}}
\end{table}
\begin{table}
\caption{Belinda brightness data (without ring subtraction)}
\resizebox{6.5in}{!}{
\begin{tabular}{|c|c|c|c|c|c|c|c|}\hline
Filename & Filter & Peak & Width1 & Width2 & Tilt & Flux & Reflectance \\
 &  & (MJy/sr) & (pixel) & (pixel) & (radian) & ($\mu$Jy) &  \\ \hline
jw02739003001\_02101\_00001\_nrcb1\_i2d.fits & F140M & 52.529 & 0.806 & 0.890 & 0.721 & 8.54 & 0.0674 \\
jw02739011001\_02101\_00004\_nrcb1\_i2d.fits & F140M & 47.696 & 0.893 & 0.919 & 0.000 & 8.87 & 0.0671 \\
jw02739011001\_02101\_00003\_nrcb1\_i2d.fits & F140M & 35.346 & 0.932 & 1.048 & 1.555 & 7.83 & 0.0592 \\
jw02739011001\_02101\_00002\_nrcb1\_i2d.fits & F140M & 56.345 & 0.897 & 0.739 & 0.022 & 8.47 & 0.0640 \\
jw02739011001\_02101\_00001\_nrcb1\_i2d.fits & F140M & 46.499 & 0.840 & 1.035 & 1.536 & 9.17 & 0.0693 \\
jw02739011001\_02101\_00005\_nrcb1\_i2d.fits & F140M & 55.206 & 1.022 & 0.732 & 0.000 & 9.37 & 0.0708 \\
\hline
\end{tabular}}
\end{table}
\begin{table}
\caption{Perdita brightness data (without ring subtraction)}
\resizebox{6.5in}{!}{
\begin{tabular}{|c|c|c|c|c|c|c|c|}\hline
Filename & Filter & Peak & Width1 & Width2 & Tilt & Flux & Reflectance \\
 &  & (MJy/sr) & (pixel) & (pixel) & (radian) & ($\mu$Jy) &  \\ \hline
jw02739003001\_02101\_00001\_nrcb1\_i2d.fits & F140M & 4.619 & 1.061 & 0.849 & 0.893 & 0.94 & 0.0612 \\
jw02739003001\_02101\_00002\_nrcb1\_i2d.fits & F140M & 5.302 & 1.059 & 0.727 & 0.164 & 0.92 & 0.0600 \\
jw02739003001\_02101\_00003\_nrcb1\_i2d.fits & F140M & 6.662 & 0.917 & 0.680 & 1.504 & 0.94 & 0.0611 \\
jw02739003001\_02101\_00004\_nrcb1\_i2d.fits & F140M & 4.924 & 0.980 & 0.794 & 0.000 & 0.87 & 0.0563 \\
jw02739011001\_02101\_00001\_nrcb1\_i2d.fits & F140M & 5.739 & 0.860 & 0.852 & 0.000 & 0.95 & 0.0592 \\
jw02739011001\_02101\_00002\_nrcb1\_i2d.fits & F140M & 4.214 & 0.903 & 0.961 & 0.439 & 0.83 & 0.0515 \\
\hline
\end{tabular}}
\end{table}

\begin{table}
\caption{Puck brightness data (without ring subtraction)}
\resizebox{6.5in}{!}{
\begin{tabular}{|c|c|c|c|c|c|c|c|}\hline
Filename & Filter & Peak & Width1 & Width2 & Tilt & Flux & Reflectance \\
 &  & (MJy/sr) & (pixel) & (pixel) & (radian) & ($\mu$Jy) &  \\ \hline
jw02739003001\_02101\_00003\_nrcb1\_i2d.fits & F140M & 195.760 & 1.009 & 0.866 & 1.314 & 38.77 & 0.0863 \\
jw02739003001\_02101\_00004\_nrcb1\_i2d.fits & F140M & 217.680 & 0.905 & 0.844 & 0.605 & 37.66 & 0.0838 \\
jw02739011001\_02101\_00001\_nrcb1\_i2d.fits & F140M & 292.877 & 0.846 & 0.724 & 0.437 & 40.63 & 0.0866 \\
jw02739011001\_02101\_00002\_nrcb1\_i2d.fits & F140M & 216.112 & 0.987 & 0.881 & 0.665 & 42.58 & 0.0907 \\
jw02739011001\_02101\_00003\_nrcb1\_i2d.fits & F140M & 191.295 & 0.946 & 0.840 & 0.653 & 34.43 & 0.0734 \\
jw02739011001\_02101\_00004\_nrcb1\_i2d.fits & F140M & 195.964 & 1.061 & 0.947 & 0.495 & 44.64 & 0.0951 \\
\hline
\end{tabular}}
\end{table}
\begin{table}
\caption{Mab brightness data (without ring subtraction)}
\resizebox{6.5in}{!}{
\begin{tabular}{|c|c|c|c|c|c|c|c|}\hline
Filename & Filter & Peak & Width1 & Width2 & Tilt & Flux & Reflectance \\
 &  & (MJy/sr) & (pixel) & (pixel) & (radian) & ($\mu$Jy) &  \\ \hline
jw02739003001\_02101\_00002\_nrcb1\_i2d.fits & F140M & 3.948 & 0.946 & 0.760 & 0.306 & 0.64 & 0.0652 \\
jw02739003001\_02101\_00003\_nrcb1\_i2d.fits & F140M & 2.863 & 1.059 & 0.944 & 0.000 & 0.65 & 0.0657 \\
jw02739011001\_02101\_00001\_nrcb1\_i2d.fits & F140M & 3.688 & 0.962 & 0.896 & 0.636 & 0.72 & 0.0699 \\
jw02739011001\_02101\_00002\_nrcb1\_i2d.fits & F140M & 4.801 & 0.922 & 0.790 & 0.000 & 0.79 & 0.0770 \\
jw02739011001\_02101\_00004\_nrcb1\_i2d.fits & F140M & 3.926 & 0.921 & 0.824 & 1.163 & 0.68 & 0.0656 \\
jw02768003001\_02103\_00005\_nrcb4\_i2d.fits & F182M & 2.052 & 1.042 & 0.879 & 0.000 & 0.43 & 0.0546 \\
\hline
\end{tabular}}
\end{table}
\begin{table}
\caption{Miranda brightness data (without ring subtraction)}
\resizebox{5.5in}{!}{
\begin{tabular}{|c|c|c|c|c|c|c|c|}\hline
Filename & Filter & Peak & Width1 & Width2 & Tilt & Flux & Reflectance \\
 &  & (MJy/sr) & (pixel) & (pixel) & (radian) & ($\mu$Jy) &  \\ \hline
jw02739003001\_02101\_00003\_nrcb1\_i2d.fits & F140M & 6406.612 & 0.750 & 0.698 & 0.103 & 760.55 & 0.1997 \\
jw02739011001\_02101\_00001\_nrcb1\_i2d.fits & F140M & 6342.454 & 0.813 & 0.807 & 0.000 & 943.66 & 0.2373 \\
jw02739011001\_02101\_00002\_nrcb1\_i2d.fits & F140M & 6567.226 & 0.738 & 0.773 & 0.000 & 848.40 & 0.2133 \\
jw02739011001\_02101\_00003\_nrcb1\_i2d.fits & F140M & 4527.247 & 0.892 & 0.882 & 0.000 & 807.66 & 0.2031 \\
jw02739011001\_02101\_00004\_nrcb1\_i2d.fits & F140M & 5374.903 & 0.903 & 0.910 & 0.000 & 1000.94 & 0.2517 \\
jw02739011001\_02101\_00005\_nrcb1\_i2d.fits & F140M & 6500.605 & 0.939 & 0.804 & 0.000 & 1112.96 & 0.2799 \\
\hline
\end{tabular}}
\end{table}

\begin{table}
\caption{Naiad brightness data (after ring subtraction)}
\resizebox{6.5in}{!}{
\begin{tabular}{|c|c|c|c|c|c|c|c|}\hline
Filename & Filter & Peak & Width1 & Width2 & Tilt & Flux & Reflectance \\
 &  & (MJy/sr) & (pixel) & (pixel) & (radian) & ($\mu$Jy) &  \\ \hline
jw02739004001\_02101\_00003\_nrcb1\_i2d.fits & F140M & 7.276 & 0.947 & 0.829 & 1.408 & 1.28 & 0.0889 \\
jw02739004001\_02101\_00005\_nrcb2\_i2d.fits & F140M & 5.676 & 1.052 & 0.896 & 0.691 & 1.23 & 0.0858 \\
jw02739004001\_02103\_00003\_nrcb1\_i2d.fits & F210M & 3.737 & 1.107 & 0.875 & 0.976 & 0.79 & 0.0850 \\
jw02739004001\_02103\_00005\_nrcb2\_i2d.fits & F210M & 3.391 & 1.205 & 1.006 & 0.311 & 0.93 & 0.0994 \\
jw02739004001\_02101\_00002\_nrcblong\_i2d.fits & F300M & 0.363 & 0.587 & 0.921 & 0.974 & 0.17 & 0.0338 \\
jw02739004001\_02101\_00003\_nrcblong\_i2d.fits & F300M & 0.226 & 0.787 & 0.931 & 0.506 & 0.14 & 0.0285 \\
jw02739004001\_02101\_00004\_nrcblong\_i2d.fits & F300M & 0.348 & 0.726 & 0.622 & 0.000 & 0.13 & 0.0270 \\
\hline
\end{tabular}}
\end{table}
\begin{table}
\caption{Thalassa brightness data (after ring subtraction)}
\resizebox{6.5in}{!}{
\begin{tabular}{|c|c|c|c|c|c|c|c|}\hline
Filename & Filter & Peak & Width1 & Width2 & Tilt & Flux & Reflectance \\
 &  & (MJy/sr) & (pixel) & (pixel) & (radian) & ($\mu$Jy) &  \\ \hline
jw02739004001\_02101\_00003\_nrcb1\_i2d.fits & F140M & 13.624 & 0.718 & 0.749 & 1.326 & 1.64 & 0.0776 \\
jw02739004001\_02101\_00005\_nrcb2\_i2d.fits & F140M & 12.664 & 0.814 & 0.779 & 1.074 & 1.85 & 0.0876 \\
jw02739004001\_02105\_00005\_nrcb2\_i2d.fits & F140M & 14.107 & 0.903 & 0.881 & 0.000 & 2.58 & 0.1224 \\
jw02739004001\_02103\_00002\_nrcb1\_i2d.fits & F210M & 6.219 & 1.063 & 0.963 & 0.000 & 1.39 & 0.1017 \\
jw02739004001\_02103\_00003\_nrcb1\_i2d.fits & F210M & 5.590 & 0.903 & 1.057 & 0.000 & 1.17 & 0.0853 \\
jw02739004001\_02107\_00003\_nrcb1\_i2d.fits & F210M & 3.821 & 1.034 & 1.533 & 0.093 & 1.33 & 0.0968 \\
jw02739004001\_02107\_00005\_nrcb2\_i2d.fits & F210M & 2.872 & 1.238 & 1.690 & 0.000 & 1.35 & 0.0988 \\
jw02739004001\_02105\_00002\_nrcblong\_i2d.fits & F300M & 0.371 & 0.852 & 0.608 & 1.526 & 0.16 & 0.0226 \\
jw02739004001\_02105\_00004\_nrcblong\_i2d.fits & F300M & 0.957 & 0.300 & 0.825 & 0.016 & 0.20 & 0.0277 \\
jw02739004001\_02105\_00005\_nrcblong\_i2d.fits & F300M & 0.342 & 0.893 & 0.592 & 1.298 & 0.15 & 0.0212 \\
\hline
\end{tabular}}
\end{table}
\begin{table}
\caption{Despina brightness data (after ring subtraction)}
\resizebox{6.5in}{!}{
\begin{tabular}{|c|c|c|c|c|c|c|c|}\hline
Filename & Filter & Peak & Width1 & Width2 & Tilt & Flux & Reflectance \\
 &  & (MJy/sr) & (pixel) & (pixel) & (radian) & ($\mu$Jy) &  \\ \hline
jw02739004001\_02101\_00001\_nrcb1\_i2d.fits & F140M & 47.153 & 0.738 & 0.935 & 0.009 & 7.28 & 0.1009 \\
jw02739004001\_02101\_00002\_nrcb1\_i2d.fits & F140M & 45.828 & 0.934 & 0.729 & 0.017 & 6.97 & 0.0966 \\
jw02739004001\_02101\_00003\_nrcb1\_i2d.fits & F140M & 43.551 & 0.917 & 0.840 & 0.295 & 7.49 & 0.1039 \\
jw02739004001\_02105\_00004\_nrcb1\_i2d.fits & F140M & 27.151 & 0.777 & 1.407 & 0.149 & 6.64 & 0.0920 \\
jw02739004001\_02105\_00003\_nrcb1\_i2d.fits & F140M & 21.989 & 0.901 & 1.325 & 0.217 & 5.86 & 0.0812 \\
jw02739004001\_02105\_00002\_nrcb1\_i2d.fits & F140M & 30.279 & 0.925 & 1.195 & 0.536 & 7.48 & 0.1037 \\
\hline
\end{tabular}}
\end{table}
\begin{table}
\caption{Galatea brightness data (after ring subtraction)}
\resizebox{6.5in}{!}{
\begin{tabular}{|c|c|c|c|c|c|c|c|}\hline
Filename & Filter & Peak & Width1 & Width2 & Tilt & Flux & Reflectance \\
 &  & (MJy/sr) & (pixel) & (pixel) & (radian) & ($\mu$Jy) &  \\ \hline
jw02739004001\_02101\_00003\_nrcb1\_i2d.fits & F140M & 46.979 & 0.941 & 0.898 & 0.311 & 8.87 & 0.1079 \\
jw02739004001\_02101\_00005\_nrcb2\_i2d.fits & F140M & 41.545 & 0.922 & 0.883 & 0.223 & 7.78 & 0.0946 \\
jw02739004001\_02103\_00003\_nrcb1\_i2d.fits & F210M & 23.849 & 1.071 & 1.089 & 0.081 & 6.09 & 0.1140 \\
jw02739004001\_02103\_00005\_nrcb2\_i2d.fits & F210M & 22.134 & 1.105 & 1.053 & 1.383 & 5.80 & 0.1086 \\
jw02739004001\_02101\_00001\_nrcblong\_i2d.fits & F300M & 1.954 & 0.739 & 1.005 & 0.142 & 1.23 & 0.0436 \\
jw02739004001\_02101\_00002\_nrcblong\_i2d.fits & F300M & 1.850 & 0.808 & 1.029 & 0.323 & 1.30 & 0.0462 \\
jw02739004001\_02101\_00003\_nrcblong\_i2d.fits & F300M & 2.082 & 0.955 & 0.820 & 0.251 & 1.38 & 0.0490 \\
jw02739004001\_02101\_00004\_nrcblong\_i2d.fits & F300M & 1.748 & 0.801 & 0.990 & 0.084 & 1.17 & 0.0416 \\
jw02739004001\_02105\_00001\_nrcblong\_i2d.fits & F300M & 1.759 & 1.067 & 0.759 & 1.504 & 1.21 & 0.0428 \\
jw02739004001\_02105\_00002\_nrcblong\_i2d.fits & F300M & 1.777 & 0.786 & 1.068 & 0.000 & 1.26 & 0.0448 \\
jw02739004001\_02103\_00001\_nrcblong\_i2d.fits & F460M & 1.510 & 0.872 & 1.279 & 0.041 & 1.48 & 0.1196 \\
jw02739004001\_02103\_00002\_nrcblong\_i2d.fits & F460M & 1.365 & 0.886 & 1.151 & 0.075 & 1.22 & 0.0989 \\
jw02739004001\_02103\_00003\_nrcblong\_i2d.fits & F460M & 1.508 & 0.960 & 0.889 & 0.626 & 1.13 & 0.0914 \\
jw02739004001\_02103\_00004\_nrcblong\_i2d.fits & F460M & 1.329 & 1.239 & 1.211 & 0.000 & 1.75 & 0.1417 \\
\hline
\end{tabular}}
\end{table}
\begin{table}
\caption{Larissa brightness data (after ring subtraction)}
\resizebox{6.5in}{!}{
\begin{tabular}{|c|c|c|c|c|c|c|c|}\hline
Filename & Filter & Peak & Width1 & Width2 & Tilt & Flux & Reflectance \\
 &  & (MJy/sr) & (pixel) & (pixel) & (radian) & ($\mu$Jy) &  \\ \hline
jw02739004001\_02101\_00001\_nrcb1\_i2d.fits & F140M & 56.588 & 0.970 & 0.846 & 1.409 & 10.38 & 0.0855 \\
jw02739004001\_02101\_00002\_nrcb1\_i2d.fits & F140M & 61.878 & 0.863 & 0.774 & 0.000 & 9.23 & 0.0761 \\
jw02739004001\_02101\_00003\_nrcb1\_i2d.fits & F140M & 59.400 & 0.933 & 0.891 & 0.647 & 11.02 & 0.0908 \\
jw02739004001\_02101\_00004\_nrcb1\_i2d.fits & F140M & 68.013 & 0.745 & 0.884 & 0.000 & 10.00 & 0.0824 \\
jw02739004001\_02105\_00002\_nrcb1\_i2d.fits & F140M & 31.301 & 0.987 & 1.557 & 0.000 & 10.75 & 0.0885 \\
jw02739004001\_02105\_00001\_nrcb1\_i2d.fits & F140M & 33.289 & 0.855 & 1.503 & 0.000 & 9.56 & 0.0788 \\
\hline
\end{tabular}}
\end{table}

\begin{table}
\caption{Hippocamp brightness data}
\resizebox{6.5in}{!}{
\begin{tabular}{|c|c|c|c|c|c|c|c|}\hline
Filename & Filter & Peak & Width1 & Width2 & Tilt & Flux & Reflectance \\
 &  & (MJy/sr) & (pixel) & (pixel) & (radian) & ($\mu$Jy) &  \\ \hline
jw02739004001\_02101\_00003\_nrcb1\_i2d.fits & F140M & 1.505 & 1.096 & 0.708 & 1.474 & 0.26 & 0.0685 \\
jw02739004001\_02101\_00004\_nrcb1\_i2d.fits & F140M & 1.460 & 0.877 & 1.384 & 0.000 & 0.40 & 0.1041 \\
jw02739004001\_02105\_00003\_nrcb1\_i2d.fits & F140M & 0.707 & 0.934 & 1.399 & 0.000 & 0.21 & 0.0543 \\
jw02739004001\_02103\_00003\_nrcb1\_i2d.fits & F210M & 0.665 & 1.247 & 0.522 & 0.864 & 0.09 & 0.0383 \\
jw02739004001\_02103\_00004\_nrcb1\_i2d.fits & F210M & 0.859 & 0.537 & 1.851 & 1.410 & 0.19 & 0.0755 \\
jw02739004001\_02107\_00003\_nrcb1\_i2d.fits & F210M & 0.841 & 1.296 & 0.792 & 0.521 & 0.19 & 0.0764 \\
jw02739004001\_02107\_00004\_nrcb1\_i2d.fits & F210M & 0.510 & 1.659 & 1.072 & 0.583 & 0.20 & 0.0803 \\
\hline
\end{tabular}}
\end{table}
\begin{table}
\caption{Proteus brightness data}
\resizebox{6.5in}{!}{
\begin{tabular}{|c|c|c|c|c|c|c|c|}\hline
Filename & Filter & Peak & Width1 & Width2 & Tilt & Flux & Reflectance \\
 &  & (MJy/sr) & (pixel) & (pixel) & (radian) & ($\mu$Jy) &  \\ \hline
jw02739004001\_02101\_00001\_nrcb1\_i2d.fits & F140M & 325.401 & 0.975 & 0.873 & 0.024 & 61.90 & 0.1086 \\
jw02739004001\_02101\_00002\_nrcb1\_i2d.fits & F140M & 341.437 & 0.918 & 0.740 & 1.468 & 51.85 & 0.0910 \\
jw02739004001\_02101\_00003\_nrcb1\_i2d.fits & F140M & 313.834 & 0.970 & 0.917 & 0.236 & 62.35 & 0.1094 \\
jw02739004001\_02101\_00004\_nrcb1\_i2d.fits & F140M & 389.829 & 0.850 & 0.752 & 0.114 & 55.70 & 0.0977 \\
jw02739004001\_02101\_00005\_nrcb1\_i2d.fits & F140M & 360.711 & 0.770 & 0.937 & 0.000 & 58.16 & 0.1021 \\
jw02739004001\_02105\_00001\_nrcb1\_i2d.fits & F140M & 284.801 & 0.857 & 0.985 & 1.311 & 53.73 & 0.0943 \\
\hline
\end{tabular}}
\end{table}

\end{document}